\documentstyle[eqsecnum,aps,preprint]{revtex}
\newcommand{\st}{\stackrel}
\newcommand{\lh}{\leftrightarrow}
\newcommand{\pp}{\partial}
\newcommand{\s}{\sigma}
\newcommand{\ep}{\epsilon}
\newcommand{\p}{\perp}
\newcommand{\be}{\begin{eqnarray}}
\newcommand{\e}{\end{eqnarray}}
\begin{document}
\tighten
\title{Transverse Spin in QCD: Radiative Corrections}
\author{{\bf A. Harindranath}\thanks{e-mail: hari@tnp.saha.ernet.in}, 
{\bf Asmita Mukherjee}\thanks{e-mail: asmita@tnp.saha.ernet.in} \\
Saha Institute of Nuclear Physics, 1/AF, Bidhan Nagar, 
	Calcutta 700064 India \\
{\bf Raghunath Ratabole}\thanks{e-mail: raghu@cts.iisc.ernet.in} \\
 Centre for Theoretical Studies, Indian Institute of Science \\
     Bangalore 560012 India \\}
\date{September 26, 2000}
\maketitle
\begin{abstract}
In this paper we address various issues connected with transverse spin in light
front QCD. The transverse spin
operators, in $A^+ = 0$ gauge, expressed in terms of the dynamical variables 
are explicitly interaction dependent unlike the helicity operator which is
interaction independent in the topologically trivial sector of light-front
QCD. Although it cannot be separated into an
orbital and a spin part, we have shown that there exists an interesting decomposition of the
transverse spin operator. We discuss the physical relevance of such a
decomposition. 
We perform a one loop renormalization of the full transverse spin operator in
light-front Hamiltonian perturbation theory for a dressed quark state. We
explicitly show that all the terms dependent on the center of mass momenta
get canceled in the matrix element. The entire non-vanishing 
contribution comes from the
fermion intrinsic-like part of the transverse spin operator as a result of
cancellation between the gluonic intrinsic-like and the orbital-like part of the
transverse spin operator.  
 We compare and contrast the
calculations of transverse spin and helicity of a dressed quark in
perturbation theory. 
\end{abstract}  
\vskip .2in
\section{Introduction}

From the early days of quantum field theory, it has been recognized that the
issues associated with the spin of a composite system in an arbitrary frame
are highly complex and non-trivial \cite{alfaro}. In equal-time
quantization, the problems arise because of the fact that the Pauli-Lubanski
operators, starting from which one can construct the spin operators in a
moving frame, are interaction dependent for a composite object. Further,
it is quite difficult to separate the center of mass and internal variables
which is mandatory in the calculation of spin. Due to these difficulties,
there has been rarely any attempt to study the spin of a moving
composite system in the conventional equal time formulation of even simple
field theoretic models, let alone Quantum Chromodynamics (QCD).
  
It is well known that in light-front field theory, in addition to the
Hamiltonian, two other operators that belong to the Poincare Group, namely,
$F^i$ ($i=1,2)$ are interaction dependent. This implies interaction
dependent spin operators and this complication is generally thought to be a
penalty one has to pay for working with light-front dynamics. In contrast,
the angular momentum operators in the familiar instant form of field theory
are interaction independent. It is interesting to investigate whether one
can understand better the physical origin of the interaction dependence in 
the light-front case. 

A second problem is that, together with the light-front helicity ${\cal
J}^3$, $F^i$ do not obey $SU(2)$ algebra, the commutation relations obeyed
by the spin operators of a massive particle. They obey $E(2)$ algebra,
appropriate for a massless particle. This implies that even though $F^i$
performs ``rotations" about the transverse axes, they have continuous
spectrum. It is, however, known how to solve this problem. In terms of the
rest of the Poincare generators, one knows\cite{bh} how to construct spin
operators ${\cal J}^i$ that together with the helicity ${\cal J}^3$ obey the
$SU(2)$ algebra. One observes that ${\cal J}^i$ is interaction dependent and
has a  highly nontrivial operator structure in contrast to ${\cal J}^3$.
Further, unlike ${\cal J}^3$, ${\cal J}^i$ cannot be separated into orbital
and spin parts. So far, most of the studies of the transverse spin operators in
light-front field theory are restricted to free field theory
\cite{except}. Even
in this case, the operators have a highly complicated structure. However,
one can write these operators as a sum of orbital and spin parts, which can
be achieved via a unitary transformation, called Melosh transformation
\cite{melosh}. In
interacting theory, presumably this can be achieved order by order in a
suitable expansion parameter \cite{bp} which is justifiable only in a weakly coupled
theory.

Knowledge about 
transverse rotation operators and transverse spin operators is mandatory 
for addressing issues concerning Lorentz invariance in
light-front theory. Unfortunately,
very little is
known \cite{review} regarding the field theoretic aspects of the
interaction dependent spin operators. 
 {\it We emphasize that in a moving frame, the spin
operators are interaction dependent irrespective of whether one considers
equal-time field theory or light-front field theory}. 
To the best of our knowledge, in gauge field theory, the canonical 
structure of spin
operators of a composite system in an {\it arbitrary} frame  has 
never been studied.

Recently it was shown that \cite{hk}, starting from the manifestly gauge 
invariant
symmetric energy momentum tensor, in light-front QCD (the gauge $A^+=0$ and
light-front variables), after the elimination of constrained variables,
${\cal J}^3$ becomes explicitly interaction independent and can be separated
into quark and gluon orbital and spin operators. Thus one can write down a
helicity sum rule which has a clear physical meaning. The orbital and
intrinsic parts of the light-front helicity operator have also been analyzed
recently in \cite{ang}. Even though ${\cal
J}^i$ cannot be separated into orbital and spin parts and they are
interaction dependent, one can still ask whether one can identify distinct 
operator structures in ${\cal J}^i$ and whether one can propose a physically
interesting decomposition. Is this decomposition protected by radiative corrections? If
distinct operators indeed emerge, do they have any phenomenological
consequences especially in deep inelastic scattering which is a light cone
dominated process?

Another important issue concerns renormalization. In light-front QCD
Hamiltonian, quark mass appears as $m^2$ and $m$ terms, $m^2$  in the
free helicity non-flip part of the Hamiltonian and $m$  in the
interaction dependent helicity flip part of the Hamiltonian. It is known
that $m^2$ and $m$ renormalize differently. $m^2$ and $m$ also appear
in ${\cal J}^i$. Do they undergo renormalization? Since ${\cal J}^i$ are
interaction dependent, do they require new counterterms in addition to those
necessary to renormalize the Hamiltonian?

In order to resolve the above mentioned problems and puzzles, we have
undertaken an  investigation of the spin of a composite
system in an arbitrary reference frame in QCD. We have compared and
contrasted both the instant form and front form formulations. In instant
form, even though the angular momentum operators are interaction
independent, they qualify as spin operators only in the rest frame of the
system. In an arbitrary reference frame, the appropriate spin operators
involve, in addition to angular momentum operators, also interaction dependent
boost operators. Thus one puzzle is resolved, namely, the interaction
dependence of the spin of a composite system in an arbitrary reference frame
is not a peculiarity of light-front dynamics, it is a general feature in any
formulation of quantum field theory. What is peculiar to light-front
dynamics is that one can at most go only to the transverse rest frame of the
particle. No frame exists in which $P^+=0$ and one is so to speak $``$always in
a moving frame". As a consequence, spin measured in any direction other than
that of $P^+$ cannot be separated into orbital and intrinsic parts. This is
to to be contrasted with the light-front helicity ${\cal J}^3$ which is independent
of interactions and further can be separated in to orbital and intrinsic
parts. The situation is quite analogous to that of a light-like particle. In
this case it is well known that since there is no rest frame, one can
uniquely identify the spin of the particle only along the direction of motion
since only along this direction one can disentangle rotation from translation
for a massless particle. Also, in any direction other than the direction of
motion, one cannot separate the angular momentum into orbital and intrinsic 
parts. 

In our earlier paper \cite{let}, we have shown that even though ${\cal J}^i$ cannot be
separated into orbital and intrinsic parts, one can still achieve a
separation into three distinct operator structures. Specifically, 
starting from the manifestly gauge invariant symmetric energy momentum
tensor in QCD, we have derived expressions for the interaction dependent 
transverse spin operators ${\cal J}^i$ ($i=1,2$) which are responsible 
for the helicity flip of the nucleon in light-front quantization. In order 
to construct ${\cal J}^i$, first we have derived expressions for the 
transverse rotation operators $F^i$. In the gauge  $A^+=0$, we eliminated the 
constrained variables. In the completely gauge fixed sector, in terms of 
the dynamical variables, we have shown that one can decompose
${\cal J}^i= {\cal J}^i_I + {\cal J}^i_{II} + {\cal J}^i_{III}$ where only 
${\cal J}^i_{I}$ has explicit coordinate ($x^-, x^i$) dependence in its 
integrand. The operators ${\cal J}^i_{II}$ and ${\cal J}^i_{III}$ arise from 
the fermionic and bosonic parts respectively of the gauge invariant energy 
momentum tensor. ${\cal J}^i_I$ is orbital-like and ${\cal J}^i_{II}$ and 
${\cal J}^i_{III}$ are fermion intrinsic-like and gluon intrinsic-like spin
operators respectively.

In this work, we explore the theoretical consequences of the decomposition of 
${\cal J}^i$. We compare and contrast the
consequences of this decomposition  and the corresponding decomposition of
the helicity operator into orbital and spin parts.
Next we address the issue of radiative corrections  by
carrying out the calculation of the transverse spin of a dressed quark in
pQCD in the old-fashioned Hamiltonian formalism. To the best of our
knowledge, this is for the first time that such a calculation has been
performed in quantum field theory. This calculation is
facilitated by the fact that boost is kinematical in the light-front
formalism. Thus we are able to isolate the internal motion which is only
physically relevant  from the spurious center of mass motion. We carry out the
calculations in a reference frame with arbitrary transverse momentum
$P^\perp$ and explicitly verify the frame independence of our results.
 We find that
because of cancellation between various interaction independent and dependent 
operator matrix elements, only one counterterm is needed. We establish the
fact the mass counterterm for the renormalization of ${\cal J}^i$ is the
same mass counterterm  required for the linear mass term appearing in
the interaction dependent helicity flip vertex in QCD. It is important to
mention that the divergence structure and renormalization in light-front
theory is entirely different from the usual equal-time theory.
If one uses constituent momentum cutoff, one violates boost invariance and also 
encounters non-analytic behavior in the structure of counterterms
\cite{wilson}.
In this paper, we have done one loop renormalization of the transverse spin
operators by imposing cutoff on the relative transverse momenta and on the
longitudinal momentum fraction.
 Upto one loop, we find that all infrared divergences (in the
longitudinal momentum fraction) get canceled in the result. The
renormalization of these operators using similarity renormalization
technique \cite{wilson} is to be done in future.

The plan of the paper is as follows. In Sec. II, first, we briefly
review the complexities associated with the description of the spin of a
composite system in a moving frame in the conventional equal time
quantization. Then we give the
explicit form of transverse rotation operators in light-front QCD.
In section III, we discuss the physical relevance of the decomposition of
the transverse spin operator and also compare and contrast it with the
helicity operator. In 
section IV, we present the calculation of the transverse spin for a dressed
quark state upto $O(\alpha_s)$ in perturbation theory. 
Discussion and conclusions are given in section V. The explicit forms of the
kinematical operators and the Hamiltonian in light-front QCD starting from
the gauge invariant symmetric interaction dependent energy momentum tensor
are derived in appendix A.  The evaluation of the transverse spin of a system of two
free fermions is given in the appendix B. The detailed derivation of the
transverse rotation operators in QCD, which are needed for the construction
of the transverse spin operators, is given in appendix C. The full evaluation of
the transverse spin operator for a dressed quark in an arbitrary reference
frame is given in appendix D. There we also show the manifest cancellation
of all the center of mass momentum dependent terms. Some details of the
calculation are provided in appendix E.

\section{The transverse spin operators in QCD}
In this section we first discuss the complexities associated with the spin
operators for a composite system in equal-time formulation and also compare
with the light-front case. Then we give the expressions for interaction dependent
transverse rotation operators in light-front QCD starting from the
manifestly gauge invariant energy momentum tensor.

The angular momentum density
\begin{eqnarray}
{\cal M}^{\alpha \mu \nu} = x^\mu \Theta^{\alpha \nu} - x^\nu
\Theta^{\alpha \mu}. 
\end{eqnarray}
In equal time theory, generalized angular momentum
\begin{eqnarray}
M^{\mu \nu} = \int d^3x {\cal M}^{0 \mu \nu}.
\end{eqnarray}
The rotation operators are $ J^i = \epsilon^{ijk} M^{jk}$. Thus in a
non-gauge theory, all the three components of the rotation operators are
manifestly interaction independent. However, the spin operators $S^i$ for a 
composite system in
a moving frame involves, in addition to  $J^i$, the  boost operators
$K^i = M^{0i}$ which are interaction dependent. 
Intrinsic spin operators in an arbitrary reference frame in equal-time 
quantization are given\cite{gur} in 
terms of the Poincare generators by,
\begin{eqnarray}
{\bf S}=&&{1\over M}\left[{\bf W}-{{\bf P}W^{0}\over {M+H}}\right]\nonumber\\
  &&={\bf J}~{P^0 \over M} - {\bf K} \times  {{\bf P}\over M} -
{({\bf J} \cdot {\bf P})\over {M+P^0}}{{\bf P}\over M}    
\end{eqnarray} 
where ${\bf W}$ are the space components of the Pauli-Lubanski
operator, $W^\mu=-{1\over 2}\epsilon^{\mu \nu \rho \lambda}M_{\nu \rho}
P_\lambda$. $ H $, ${\vec P}$ are equal time Hamiltonian and momentum
operators respectively obtained by integrating the energy
momentum tensor over a spacelike surface and $ {\vec J}$ and $ {\vec K}$ are
the equal time rotation and boost generators respectively, which are obtained by
integrating the angular momentum density over a spacelike surface.  
Since boost ${\bf K}$ is dynamical, {\it all the three components of ${\bf S}$
are interaction dependent} in the equal time quantization. 
Nevertheless, the component of {\bf S} along {\bf P} remains kinematical.
This is to be
compared with light-front quantization where {\it the third component of the
light-front spin operator ${\cal J}^3$ is kinematical}.
This arises from the facts that boost operators are kinematical on the light
front, the interaction dependence of light-front spin
operators ${\cal J}^i$ arises solely from the rotation operators, and the
third component of the rotation operator $J^3$ is kinematical on the light
front.

A further  complication arises in equal time quantization. In order
to describe the intrinsic spin of a composite system, one should be able to
separate the center of mass motion from the internal motion. Even in free field
theory, this turns out to be quite involved (See Ref. \cite{os} and references
therein). On the other hand, in light-front theory, since transverse boosts
are simply Galilean boosts, separation of center of mass motion and internal
motion is as simple as in non-relativistic theory (see appendix D).

A gauge invariant separation of the nucleon angular momentum is performed in
Ref. \cite{ji}. However, as far the spin operator in an arbitrary reference
frame is concerned, 
the analysis of this reference is valid only in the rest frame where spin
coincides with total angular momentum operator. Further, there is no mention
of the complications in the equal time theory, which arise from 
 the need to project out the
center of mass motion in an arbitrary reference frame. 
Moreover, the distinction between the longitudinal and transverse components
of the spin is not made.
It is crucial to make this distinction since physically
the longitudinal and transverse components of the spin carry quite distinct
information (as is clear, for example, from the spin of a massless particle). 
Moreover, even for the third component of the spin of a composite system
in a moving frame, there is crucial difference between equal time and light
front cases. ${\cal J}^3$
(helicity) is interaction independent whereas $S^3$ is interaction
dependent in general except when measured along the direction of {\bf P}.

In light-front theory, generalized angular momentum 
\begin{eqnarray}
M^{\mu \nu} =
{ 1 \over 2} \int dx^- d^2 x^\perp {\cal M}^{+ \mu \nu}.
\end{eqnarray}
$J^3$ which is related to the helicity is given by  
\begin{eqnarray}
J^3 = M^{12} = { 1
\over 2} \int dx^1 d^2 x^\perp [x^1
\Theta^{+2} - x^2 \Theta^{+1} ] 
\end{eqnarray}
and is interaction independent.
On the other hand, the transverse rotation operators which are related
to the transverse spin are given by
$$ F^i =M^{-i}= { 1 \over 2} \int dx^- d^2 x^\perp [ x^- \Theta^{+i} - x^i
\Theta^{+-} ] . $$
They are interaction dependent {even in a non-gauge theory} since
$\Theta^{+-}$ is the Hamiltonian density.

For a massive particle, the transverse spin operators\cite{bh} ${\cal J}^i$ in 
light-front theory are given in terms of Poincare generators by
\begin{eqnarray}
M{\cal J}^1 &&= W^1 - P^1 {\cal J}^3 = { 1 \over 2} F^2 P^+ + K^3 P^2   - { 1
\over 2} E^2 P^- - P^1 {\cal J}^3, \label{j1}
\e
\be  
M{\cal J}^2 &&= W^2 - P^2 {\cal J}^3 = - { 1 \over 2 } F^1 P^+ -K^3 P^1  + { 1 \over 2} E^1 P^- -
P^2 {\cal J}^3. \label{j2}
\end{eqnarray}
The first term in Eqs. (\ref{j1}) and (\ref{j2}) contains both center of
mass motion and internal motion and the next three terms in these equations
serve to remove the center of mass motion.  
 
The helicity operator is given by
\begin{eqnarray}
{\cal J}^3 &&= {W^+ \over P^+} = J^3 + { 1 \over P^+}(E^1P^2 - E^2 P^1).
\end{eqnarray} 
Here, $J^3$ contain both center of mass motion and internal motion and the
other two terms serve to remove the center of mass motion. 
The operators ${\cal J}^i$  obey the angular momentum commutation relations 
\begin{eqnarray}
\left [ {\cal J}^i, {\cal J}^j \right ] = i \epsilon^{ijk} {\cal J}^k
\label{j3} .
\end{eqnarray}

In order to calculate the transverse spin operators, first we need to
construct the Poincare generators $P^+$, $P^i$, $P^-$, 
$K^3$, $E^i$, $J^3$ and
$F^i$ in light-front QCD. The explicit form of the operator $J^3$ is given
Ref. \cite{hk}. The construction of $F^i$ which is algebraically 
quite involved is carried out in appendix C. The final form of $F^i$ is also
given in Ref.\cite{let}. The construction of the
rest of the kinematical operators is given in Appendix A. In this appendix
we have presented all the operators in the manifestly Hermitian form, which
is necessary, as we shall see later.

In order to have a physical picture of the complicated situation at hand, it
is instructive to calculate the spin operator in free 
field
theory. The case of two free massive fermions is carried out in Appendix B.

In light-front theory we set the gauge $A^+=0$
and eliminate the dependent variables $\psi^-$ and $A^-$ using the equations
of constraint. We have shown that \cite{let} (for details of the derivation
see appendix C), in the topologically trivial sector
of the theory one can write the transverse rotation operator as,
\begin{eqnarray}
F^2  = F^2_{I} + F^2_{II} + F^2_{III},
\end{eqnarray}
where
\begin{eqnarray}
F^2_{I}&& = {1\over 2} \int dx^- d^2x^\p [ x^- {\cal P}^2_0 - x^2 ({\cal H}_0 +
{\cal V}) ], \label{f21}\\
F^2_{II} &&= 
{1\over 2} \int dx^- d^2x^\p \Bigg [\xi^\dagger \Big [ \sigma^3 \partial^1 + i \partial^2
\Big]{ 1 \over
\partial^+} \xi + \Big [ { 1 \over \partial^+} (\partial^1 \xi^\dagger \sigma^3 -
i \partial^2 \xi^\dagger) \Big ] \xi \Bigg ] \nonumber \\ 
&&~~~~~~ + {1\over 2} \int dx^- d^2x^\p m \Bigg [ \xi^\dagger \Big [{ \sigma^1 \over i \partial^+} 
\xi\Big ] -
\Big [{ 1 \over i \partial^+} \xi^\dagger\sigma^1\Big ] \xi \Bigg ]
\nonumber \\
&& ~~+ {1\over 2} \int dx^- d^2x^\p  g \Bigg [ \xi^\dagger { 1 \over
\partial^+}[(-i \sigma^3 A^1 + A^2)\xi] + { 1 \over \partial^+}
[ \xi^\dagger (i \sigma^3 A^1 + A^2)]\xi \Bigg ], \\
F^2_{III}&&= 
- \int dx^- d^2 x^\perp 2 (\partial^1 A^{1})A^2 \nonumber \\
&&~~-{1\over 2} \int dx^- d^2x^\p g {4\over {\pp^+}} (\xi^\dagger T^a
\xi) A^{2a} - {1\over 2} \int dx^- d^2x^\p g f^{abc} {2\over {\pp^+}} (
A^{ib} \pp^+ A^{ic} ) A^{2a}.
\end{eqnarray}
Here $ {\cal P}^i_0$ is the free momentum density, $ {\cal H}_o$ is the
free Hamiltonian density and ${\cal V}$ are the interaction terms in the
Hamiltonian in manifestly Hermitian form (see Appendix A). The operators $F^2_{II}$ and $F^2_{III}$ 
whose integrands do not
explicitly depend upon coordinates arise from the fermionic and bosonic
parts respectively of the gauge invariant symmetric energy momentum tensor
in QCD. The above separation is slightly different from that in \cite{let}. 
From Eq. (\ref{j1}) in Sec. II it follows that the transverse spin
operators ${\cal J}^i$, ($i=1,2$) can also be written as the sum of three
parts, ${\cal J}^i_{I}$ whose integrand 
has explicit coordinate dependence, ${\cal
J}^i_{II}$ which arises from the fermionic part, and  ${\cal J}^i_{III}$ which
arises from the bosonic part of the energy momentum tensor.   

In the next section, we propose a decomposition of transverse spin in
analogy with the helicity case and compare and contrast the two cases.
\section{the decomposition of transverse spin}
The transverse spin operators ${\cal J}^i$ in light-front theory for a
massive particle can be given in terms of Poincare generators by
Eq.(\ref{j1}). 
In \cite{hk} it has been shown explicitly 
that the helicity operator ${\cal J}^3$ in the light-front gauge, 
in terms of the dynamical fields in the topologically
trivial sector of QCD can be written as,
\be
{\cal J}^3 = {\cal J}_{fi}^3 + {\cal J}_{fo}^3 +{\cal J}_{go}^3 +{\cal J}_{
gi}^3
\e
where  ${\cal J}_{fi}^3$ is the fermion intrinsic part, ${\cal J}_{fo}^3$ 
is the fermion orbital part, ${\cal J}_{go}^3$ is the gluon orbital part and 
${\cal J}_{gi}^3$ is the gluon intrinsic part. The helicity sum rule is
given by, for a longitudinally polarized fermion state,
\be
{1\over {\cal N}} \langle P S^\parallel \mid {\cal J}_{fi}^3 + {\cal J}_{fo}^3 +{\cal J}_{go}^3 +{\cal J}_{
gi}^3\mid P S^\parallel \rangle = \pm {1\over 2}.
\e
In the transverse rest frame ($P^\perp=0$), the helicity sum rule takes the
form,
\be
{1\over {\cal N}} \langle P S^\parallel \mid J_{fi}^3 + J_{fo}^3 +
J_{go}^3 +J_{
gi}^3\mid P S^\parallel \rangle = \pm {1\over 2}.
\e

For a boson state, RHS of the above equation should be replaced with the
corresponding helicity. Here, ${\cal N}$ is the normalization constant of
the state. 
Unlike the helicity operator, which can be
separated into orbital and spin parts, the transverse spin operators cannot
be written as a sum of orbital and spin contributions. Only in the free theory, 
one can write them as a sum of orbital and spin parts by a unitary
transformation called Melosh transformation. However, we have
shown that they can be separated into three distinct components. At this
point, we would also like to contrast our work with Ref.\cite{ji}, 
where a gauge
invariant decomposition of nucleon spin has been done. The analysis in
Ref.\cite{ji} has been performed in the rest frame of the hadron and no distinction
is made between  helicity and transverse spin, whereas, we have worked in the
gauge fixed theory in an arbitrary reference frame. 

In analogy with the helicity sum rule, we propose a decomposition of the
transverse spin, which can be written as, 
\be
{1\over {\cal N}} \langle P S^\perp \mid {\cal J}^i_I + {\cal J}^i_{II}+ {\cal
J}^i_{III} \mid P S^\perp \rangle = \pm {1\over 2}
\label{tsum}
\e
for a fermion state polarized in the transverse direction. For a bosonic state, RHS will be replaced with the
corresponding transverse component of spin.

What is the physical relevance of such a
decomposition of the transverse spin operator?  The fermion intrinsic part of the
helicity operator can be
related to the first moment of the quark helicity distribution
measured in longitudinally polarized deep inelastic scattering. In the case
of the transverse spin operator, we have shown\cite{let} that there exists 
a direct connection between
the hadron expectation value of the fermionic intrinsic-like part of the transverse spin
operator ${\cal J}^i_{II}$ and the integral of the quark distribution 
function $g_T$ that
appear in transversely polarized deep inelastic scattering. Also we can
identify\cite{let} the operators that are present in the hadron expectation value
of  
${\cal J}^i_{III}$ with the operator structures that are present in 
the integral of the gluon distribution function that appear in transverse
polarized hard scattering. 
The physical relevance of the
decomposition is made clear from the identification. Our results show 
 the intimate
connection between transverse spin in light-front QCD and transverse
polarized deep inelastic scattering. As far as we know, such connections are
not established so far  in instant form of field theory and this is the
first time that the first moment of $g_T$ is related to a conserved
quantity. It is already known
that the interaction independent light-front helicity operator
${\cal J}^3$ can be separated as ${\cal J}^3 = {\cal J}^3_{q(i)} +
{\cal J}^3_{q(o)} + {\cal J}^3_{g(i)} + {\cal J}^3_{g(o)}$  and further, 
hadron expectation value of ${\cal J}^3_{q(i)}$ is directly related to the
integral of the deep inelastic helicity structure function $g_1$. Thus we
find natural physical explanation for the simplicity and complexity of
operator structures appearing in the structure functions $g_1$ and $g_T$
respectively. Another
important point is that in perturbation theory, 
the helicity flip interactions which are proportional
to mass play a crucial role both in $g_T$ and in the transverse spin
operator whereas they are not important in the case of the helicity
operator.

Because the transverse spin operators are interaction dependent, they
acquire divergences in perturbation theory. One has to regularize them by
imposing momentum cutoffs and
in the regularized theory the Poincare algebra as well as the commutation 
relation obeyed by the spin operators are violated \cite{glazek}. One has to
introduce appropriate counterterms to restore the algebra. 
In the next section, we perform the renormalization of the full
transverse spin operator upto $O(\alpha_s)$ in light-front Hamiltonian
perturbation theory by evaluating the matrix element for a quark state
dressed with one gluon. This calculation also verifies the relation
(\ref{tsum})  
 upto $O(\alpha_s)$ in perturbation theory.

\section{Transverse spin of a dressed quark in perturbation theory}

In this section, we evaluate the expectation value of the transverse spin
operator in perturbative QCD for a dressed quark state.

The dressed quark state with fixed helicity $\sigma$ can be expanded in Fock space as,
\be
\mid P, \sigma \rangle && = \phi^\lambda_1 b^\dagger(P,\sigma) \mid 0 \rangle
\nonumber \\  
&& + \sum_{\sigma_1,\lambda_2} \int 
{dk_1^+ d^2k_1^\perp \over \sqrt{2 (2 \pi)^3 k_1^+}}  
\int 
{dk_2^+ d^2k_2^\perp \over \sqrt{2 (2 \pi)^3 k_2^+}}  
\sqrt{2 (2 \pi)^3 P^+} \delta^3(P-k_1-k_2) \nonumber \\
&& ~~~~~\phi^\sigma_{\sigma_1, \lambda_2}(P,\mid k_1,; k_2 ) b^\dagger(k_1,
\sigma_1) a^\dagger(k_2, \lambda_2) \mid 0 \rangle. 
\label{dr0}
\e 
We are considering  dressing with one gluon since we shall evaluate
the expectation value upto $O(g^2)$. The normalization of the state is given by,
\be
\langle k',\lambda' \mid k,\lambda \rangle = 2(2\pi)^3 k^+ \delta_{\lambda
\lambda'} \delta(k^+ -k'^+) \delta ( k^\perp - k'^\perp).
\label{nor}
\e  
The quark target transversely polarized in the $x$ direction can be expressed
in terms of helicity up and down states by,
\be
\mid k^+, k^\perp, s^1 \rangle = {1\over {\sqrt 2}}(\mid k^+, k^\perp,
\uparrow \rangle \pm \mid k^+, k^\perp, \downarrow \rangle)
\label{dr1}
\e 
with $s^1 = \pm m_R$, where $m_R$ is the renormalized mass of the quark.

We introduce the boost invariant amplitudes $\Phi_1^\lambda$ and
$\Phi^\lambda_{\sigma_1 \lambda_2}(x, q^\perp)$ respectively by
$\phi^\lambda (k) = \Phi_1^\lambda$ and $\phi^\lambda_{\lambda_1 \lambda_2}
(k; k_1, k_2)= {1\over {\sqrt k^+}}\Phi^\lambda_{\lambda_1 \lambda_2}(x,
q^\perp)$, where $x={k_1^+ \over P^+}$ and $q^\perp = k_1^\perp -
x P^\perp$ . From the light-front QCD Hamiltonian, to lowest order in
perturbative QCD, we have,
\be
	&& \Phi_{\sigma_1, \sigma_2}^\lambda (x,q^\perp) 
		= -{ x(1-x) \over { (q^\perp)^2 + m^2(1-x)^2}}{1\over
{\sqrt {1-x}}} \nonumber \\
	&&~~~~~~~~~~\times   { g \over {\sqrt {2 (2 \pi)^3}}} T^a 
		\chi^\dagger_{\sigma_1} \Big[  2 {q^\perp \over 1-x} + 
		{{\tilde \sigma^\perp.q^\perp}\over x} \tilde \sigma^\perp - 
		\tilde \sigma^\perp i m {{1-x}\over x} \Big ] \chi_{\lambda} .{(\epsilon^\perp_{
		\sigma_2})}^*\Phi^\lambda_1. \label{dr2} 
\label{wf}
\e 
Here $m$ is the quark mass and $x$ is the longitudinal momentum carried by
the quark. Also, $\tilde \sigma^1 = \sigma^2$ and $ \tilde \sigma^2 = -
\sigma^1$. It is to
be noted that the $m$ dependence in the above wave function arises from
the helicity flip part of the light-front QCD Hamiltonian. This term plays a
very important role in the case of transversely polarized target states.

For simplicity, in this section, we calculate the matrix element of the transverse spin
operator for a dressed quark state in a frame where the transverse momentum of
the quark is zero. It can be seen from Eq. (\ref{j1}) that the sole contribution
in this case comes from the first term in the RHS, namely the
transverse rotation operator. A detailed calculation of the matrix elements
of the transverse spin operator in an arbitrary reference frame is given in
appendix D where we have explicitly shown that all the terms depending on
$P^\perp$ get canceled.

The matrix elements presented below have been evaluated between 
states of different helicities, namely $\sigma$ and
$\sigma'$. Since the transversely polarized  state can be
expressed in terms of the longitudinally polarized (helicity) states by
Eq. (\ref{dr1}), the matrix elements of these operators between transversely
polarized states can be easily obtained from these expressions.   

Here, we have used the manifest Hermitian form of all the operators. It is
necessary to keep manifest Hermiticity at each intermediate step to cancel
terms containing derivative of delta function.
 
The operator ${1\over 2}F^2P^+$ can be separated into three parts,
\be
{1\over 2}F^2P^+ = {1\over 2}F^2_IP^+ + {1\over 2}F^2_{II}P^+ +
 {1\over 2}F^2_{III}P^+,
\e
where $F^2_I$, $F^2_{II}$ and $F^2_{III}$ have been defined earlier.
The matrix elements of the different parts of these for a dressed quark state
are given below. The evaluation of the matrix element of ${1\over 2}F^2_IP^+$
 is quite complicated since it involves derivatives of delta functions. A
part of this calculation has been given in some detail in appendix E.
The operator,
\be
{1\over 2}F^2_IP^+ = {1\over 2}F^2_I(1)P^+ - {1\over 2}F^2_I(2)P^+ -
 {1\over 2}F^2_I(3)P^+.
\label{f2i}
\e
The first term contains the momentum density, the second and the third terms
contain the free and the interaction parts of the Hamiltonian density respectively.
The matrix elements are given by,
\be
\langle P,\sigma \mid {1\over 2}F^2_I(1)P^+ \mid P, \sigma' \rangle  &&=
\langle P,\sigma \mid {1\over 2} \int dx d^2q^\perp x^- P^2_0 {1\over
2}P^+ \mid P, \sigma' \rangle \nonumber\\&& = -{i\over 2}\sum_{spin} 
\int dx d^2q^\perp q^2 
\Phi^{*\sigma}_{\sigma_1\lambda}{\pp\Phi^{*\sigma'}_{\sigma'_1\lambda'} \over
{\pp x}} + h. c. 
\e
\be
\langle P,\sigma \mid && {1\over 2}F^2_I(2)P^+\mid P, \sigma' \rangle = 
\langle P,\sigma \mid {1\over 2} \int dx d^2q^\perp x^2 P^-_0 {1\over
2}P^+\mid P, \sigma' \rangle \nonumber\\&& ={i\over 4}\sum_{spin}\int dxd^2q^\perp 
\Phi^{*\sigma}_{\sigma_1\lambda}{\pp \Phi^{\sigma'}_{\sigma'_1\lambda'}
\over {\pp q^2}} (q^\perp)^2 \left( {{1-x}\over x} - {x\over {1-x}} \right ) 
\nonumber\\&&~~~~~~~~~ +{i\over 4}\sum_{spin} \int dx d^2q^\perp m^2 {{1-x}\over x}
\Phi^{*\sigma}_{\sigma_1
\lambda}{\pp \Phi^{\sigma'}_{\sigma'_1\lambda'}\over {\pp q^2}}+ h. c. 
\e
In the above two equations, both the single particle and two particle
diagonal matrix elements contribute. Here, {\it h.c.} is the Hermitian conjugate, $\sum_{spin}$ is the summation over $\sigma_1, {\sigma'}_1, \lambda_1,
{\lambda'}_1$. $P^-_0$ is the free part of the Hamiltonian density.
\be
\langle P, \sigma \mid &&{1\over 2}F^2_I(3)P^+\mid P, \sigma' \rangle =
\langle P, \sigma \mid  {1\over 2} \int dx d^2q^\perp x^2 P^-_{int} {1\over
2}P^+\mid P, \sigma' \rangle \nonumber\\&& = {g\over \sqrt {2(2\pi)^3}}
\sum_{spin} \int dxd^2q^\perp {1\over {\sqrt {1-x}}} \Big (
-{i\over 4} 
\Phi_1^{* \sigma} \chi^\dagger_\sigma [ {\tilde \sigma}^2 ( {\tilde
\sigma}^\perp \cdot \ep^\perp ) \nonumber\\&&~~~~~~~~~~~~~~~~~~~+ {({\tilde 
\sigma}^\perp \cdot \ep^\perp)
{\tilde \sigma}^2 \over x}]\chi_{\sigma_1} \Phi^{\sigma'}_{\sigma_1
\lambda}+ h. c.\Big ).
\e
$P^-_{int}$ is the interaction part of the light-front QCD Hamiltonian
density. Only the $qqg$ part of it contributes to the dressed quark matrix
element.

The operator ${1\over 2}F^2_{II}P^+$ which originates from the fermionic
part of the energy momentum tensor, can be separated into three parts,

\be
{1\over 2}F^2_{II}P^+ = {1\over 2}F^2_{mII}P^+ + {1\over 2}F^2_{q^\perp II}P^++ 
{1\over 2}F^2_{gII}P^+
\e
where ${1\over 2}F^2_{mII}P^+$ is the  explicit mass
dependent part of the operator, ${1\over 2}F^2_{q^\perp II}P^+$ is the part
containing derivatives with respect to $x^\perp$ and ${1\over 2}F^2_{g II}P^+$ is
the interaction part. The matrix elements are given by, 
\be
\langle P, \sigma \mid {1\over 2}F^2_{mII}P^+ \mid P, \sigma'
\rangle = {m\over 2}\Phi^{*\sigma}_1
\Phi^{\sigma'}_1 + {m\over 2}\sum_{spin} \int dx d^2q^\perp  \Phi^{*\sigma}_{\sigma_1
\lambda} \chi^\dagger_{\sigma_1} \sigma^1 \chi_{\sigma'_1}
\Phi^{\sigma'}_{\sigma'_1
\lambda'} {1\over x}, \label{m2}
\e
\be
\langle P, \sigma \mid {1\over 2}F^2_{q^\perp II} P^+ \mid P, \sigma'
 \rangle &&= 
{1\over 2}\sum_{spin}\int dx d^2q^\perp \Phi^{*\sigma}_{\sigma_1
\lambda} \chi^\dagger_{\sigma_1} \sigma^3 q^1 \chi_{\sigma'_1}
\Phi^{\sigma'}_{\sigma'_1 \lambda'} {1\over x},
\label{q2}
\e
\be
\langle P, \sigma \mid {1\over 2}F^2_{g II} P^+ \mid P, \sigma'
 \rangle &&= {1\over 4} {g\over {\sqrt {2(2 \pi)^3}}}\sum_{spin} \int dx d^2q^\perp
{1\over {\sqrt {1-x}}} \Big ( i\Phi^{*\sigma}_1 \nonumber\\&&~~~~~~\Big [ 
\chi^\dagger_\sigma (
-i\sigma^3 \ep^1_\lambda + \ep^2_\lambda ) \chi_{\sigma_1} - {1\over x}
\chi^\dagger_\sigma (i\sigma^3 \ep^1_\lambda + \ep^2_\lambda ) \chi_{\sigma_1}
 \Big ] \Phi^{\sigma'}_{\sigma_1 \lambda}  + h. c.\Big ) . \label{g2}
\e
In Eqs. (\ref{m2}) and (\ref{q2}), contributions come from only diagonal
matrix elements whereas Eq. (\ref{g2}) contain only off-diagonal matrix
elements.
The matrix element of ${1\over 2}F^2_{III} P^+$, which comes from the
gluonic part, is given by,
\be
\langle P, \sigma \mid {1\over 2}F^2_{III} P^+&& \mid P, \sigma'
 \rangle = -{g \over {\sqrt {2(2\pi)^3}}}
\sum_{spin} \int dx d^2q^\perp {1\over
{\sqrt {1-x}}}\nonumber\\&&~~~ \Big (  \Phi^{*\sigma}_1 \ep^2_\lambda 
\Phi^{\sigma'}_{\sigma_1 \lambda}{1\over {i(1-x)}} + h. c. \Big )
 - \int dx d^2q^\perp {q^1\over (1-x)} \sum_{\lambda,
\sigma_1, \sigma'_1} \lambda
\Phi^{*\sigma}_{\sigma_1 \lambda}\Phi^{\sigma'}_{\sigma'_1
\lambda} .
\e
The first term in the RHS is the off-diagonal contribution which comes from
the interaction dependent part of the operator. The second term is 
 the diagonal contribution coming from the free part.

The expectation value of the transverse spin operator between transversely
polarized states is given by,
\be
\langle P, S^1 \mid M{\cal J}^1 \mid P, S^1
 \rangle = \langle P,S^1 \mid {1\over 2} F^2 P^+ + K^3P^2 - {1\over 2}
E^2 P^- - P^1 {\cal J}^3 \mid P, S^1  \rangle.
\e
 Since we are in the reference frame with zero $P^\perp$, only the first term
in the RHS, i.e. the ${1\over 2} F^2 P^+$ term will contribute, as mentioned
earlier. We substitute
for $\Phi^\sigma_{\sigma_1 \lambda}$ using Eq. (\ref{wf}). The final forms of
the matrix elements are given by, 
\be
\langle P, S^1 \mid M {\cal J}^1_I(1) \mid P, S^1 \rangle = -{m
\alpha_s\over {4 \pi}}C_f ln {Q^2\over \mu^2} \int_\ep^{1-\ep} dx (1+x),
\e
\be
\langle P, S^1 \mid M {\cal J}^1_I(2) \mid P, S^1 \rangle = {m
\alpha_s\over {4 \pi}}C_f ln {Q^2\over \mu^2} \int_\ep^{1-\ep} dx (1-2x) ,
\e
\be
\langle P, S^1 \mid M {\cal J}^1_I(3) \mid P, S^1 \rangle = -{m
\alpha_s\over {4 \pi}}C_f ln {Q^2\over \mu^2} \int_\ep^{1-\ep} dx (1-x)
\e
where $M{\cal J}^1_I(1), M{\cal J}^1_I(2)$ and $ M{\cal J}^1_I(3)$ are
 related respectively to $F^2_I(1), F^2_I(2)$ and $ F^2_I(3)$ defined
earlier. 
 $\mu $ is the hadronic factorization scale for separating the 'hard'
and 'soft' dynamics of QCD, i. e. we have set a hadronic scale such that
${\mid q^\perp \mid }^2 >> \mu^2 >>m^2$. $\ep$ is a small cutoff on the
longitudinal momentum fraction.

So we obtain, from the above three expressions, using
Eq. (\ref{f2i}),
\be
\langle P, S^1 \mid M {\cal J}^1_I \mid P, S^1 \rangle = -{m
\alpha_s\over {4 \pi}}C_f ln {Q^2\over \mu^2} .
\label{dr4}
\e
The contribution to the matrix element of $M{\cal J}^1_{II}$ entirely comes from
$F^2_{II}$. The various parts of this matrix element are given by,
\be
\langle P, S^1 \mid M {\cal J}^1_{mII} \mid P, S^1 \rangle = {1\over 2} m
{\mid \Phi^\sigma_1 \mid }^2 +
{m\alpha_s \over {2 \pi}} C_f ln{Q^2\over \mu^2}\int_\ep^{1-\ep} dx{1\over {1-x}},
\label{dr6}
\e
\be
\langle P, S^1 \mid M {\cal J}^1_{q^\perp II} \mid P, S^1 \rangle = 
-{m\alpha_s \over {4\pi}} C_f ln{Q^2\over \mu^2}\int_\ep^{1-\ep} dx(1-x),
\e
\be
\langle P, S^1 \mid M {\cal J}^1_{g II} \mid P, S^1 \rangle = 
{m\alpha_s \over {4\pi}} C_f ln{Q^2\over \mu^2}{1\over 2},
\e
where $M{\cal J}^1_{m II}, M{\cal J}^1_{q^\perp II}$ and $ M{\cal J}^1_{g
II}$ are related respectively to $F^2_{mII}, F^2_{q^\perp II}$ and $ F^2_{g
II}$. In Eq. (\ref{dr6}) we have to use the normalization condition,
\be
{\mid \Phi^\sigma_1 \mid }^2 = 1- {\alpha_s\over {2\pi}} C_f ln{Q^2 \over
\mu^2} \int_\ep^{1-\ep}dx {1+x^2\over {1-x}}.
\e 
Upto $O(\alpha_s)$, the normalization condition will contribute only in the
first term of Eq. (\ref{dr6}). 
We get, from Eq. (\ref{dr6}),
\be
\langle P, S^1 \mid M {\cal J}^1_{mII} \mid P, S^1 \rangle = {1\over 2} m
+{m\alpha_s \over {4 \pi}} C_f ln{Q^2\over \mu^2}\int_\ep^{1-\ep} dx \big (
{2\over {1-x}}-{{1+x^2}\over {1-x}} \big ),
\e
It is clear that the singularity at $x=1$ is canceled due to
the contribution from the normalization condition.
The overall contribution coming from $M{\cal J}^1_{II}$ is given by,
\be
\langle P, S^1 \mid M {\cal J}^1_{II} \mid P, S^1 \rangle = {m\over 2} \left (1+
{3\alpha_s \over {4\pi}} C_f ln{Q^2\over \mu^2}\right ),
\e
which does not involve any $x$ divergence.
 The matrix element of $M{\cal J}^1_{III}$ is given by,
\be
\langle P, S^1 \mid M {\cal J}^1_{III} \mid P, S^1 \rangle = 
{2m\alpha_s \over {4\pi}} C_f ln{Q^2\over \mu^2}\int_\ep^{1-\ep} ( 1-x) dx.
\label{dr5}
\e 
It is to be noted that all the contributing matrix elements are proportional
to the quark mass. Among the different parts of the operator, only ${\cal
J}^i_{mII}$ and a part of the interaction terms in ${\cal J}^i_I$ (see
Eq. (\ref{f21})) are proportional to the quark mass $m$. These mass dependent
terms flip the quark helicity. It is also to be noted that the terms
proportional to $m^2$ do not flip the helicity. In all the other terms, though the operators
do not depend on $m$ explicitly, the contributions to the matrix elements
arise from the interference of the $m$ terms in the wave function of
Eq. (\ref{wf}), with the non-$m$ dependent terms through the different parts
of the transverse spin operator. Since in light-front formulation, helicity
and chirality are the same, these linear in $m$ terms are explicit 
chiral symmetry
breaking terms.
From Eq. (\ref{dr4}) and Eq. (\ref{dr5}) we find that,
\be
\langle P, S^1 \mid M {\cal J}^1_{I}+ M {\cal J}^1_{III} \mid P, S^1 \rangle =  
{m \alpha_s \over {4 \pi}} C_f ln {Q^2 \over \mu^2} \int_\ep^{1-\ep} (1-2x) dx
= 0
\e
which means that the entire contribution to the matrix element of the
transverse spin operator is given by,
\be
\langle P, S^1 \mid M {\cal J}^1 \mid P, S^1 \rangle =
{m\over 2} \left ( 1 + {3 \alpha_s \over {4\pi}} C_f ln{Q^2\over \mu^2}
\right).
\e
This contribution entirely comes from $M{\cal J}^1_{II}$. Contribution from the
orbital-like part $(M{\cal J}^1_I)$ exactly cancels the contribution from the
gluon intrinsic-like part $(M{\cal J}^1_{III})$. 

The renormalized mass $m_R$ of the quark is given in terms of the bare
mass upto order $\alpha_s$ in light-front Hamiltonian perturbation theory
by\cite{hari3},
\be
m_R = m \left( 1 + { 3 \alpha_s \over {4 \pi}}C_f ln {Q^2 \over \mu^2} \right
).
\e
In the light-front formulation of QCD, there are two mass terms in the
Hamiltonian, one is quadratic in $m$ which is present in the free part and
does not break chiral symmetry, the other is linear in $m$ which we discuss
here and which explicitly cause chiral symmetry breaking. An important
feature of light-front QCD is that, these two mass scales are renormalized
differently even in the perturbative region. The renormalization of $m^2$ is
different from the result stated above. 

Adding all the parts, for a dressed quark in perturbation theory upto $O(g^2)$, the
expectation value of the transverse spin operator is given by, 
\be
\langle P, S^1 \mid M {\cal J}^1 \mid P, S^1 \rangle =&& 
\langle P, S^1 \mid M {\cal J}^1_{I}+M {\cal J}^1_{II}+M {\cal J}^1_{III} 
 \mid P, S^1\rangle \nonumber\\&& = {m_R\over 2} .
\e    
It is important to mention that here we are calculating the expectation
value of the operator $M{\cal J}^i$. In order to extract the eigenvalue of 
${\cal J}^i$ 
one has to know the eigenvalue of $M$. Both $M{\cal J}^i$ and $M$ are
dynamical operators. However, in this case, 
the mass $M$ in the LHS in the renormalized theory is nothing but the
renormalized mass of the quark, which therefore gets canceled from the
above equation, and we get,
\be
\langle P, S^1 \mid  {\cal J}^1 \mid P, S^1 \rangle =&& 
\langle P, S^1 \mid  {\cal J}^1_{I}+ {\cal J}^1_{II}+ {\cal J}^1_{III} 
 \mid P, S^1 \rangle \nonumber\\&& = {1\over 2}.
\e

The identification of ${\cal J}$ with spin, therefore, requires knowledge of
the mass eigenvalue, independently of the boost invariance properties of the
light-front dynamics.

We can explicitly verify the relation between the integral of $g_T$ and the
expectation value of the fermion intrinsic-like part of the transverse spin
operator to order $\alpha_s$ in perturbative QCD.
The transverse polarized structure function for a dressed quark is 
given\cite{hari1} by,
\be
g_T(x, Q^2) = &&{e^2_q\over 2}{m\over S^1} \Big \{ \delta (1-x) +
 {\alpha_s \over {2\pi}}  C_f ln{Q^2\over \mu^2} \Big [ {1+2x -x^2\over
 {1-x}} - \delta(1-x) \int_0^1 dx' {1+{x'}^2\over {1-x'}}
\nonumber\\&&~~~~~~~~~~~~~~~~~~~~~~~~~+{1\over 2}
\delta (1-x) \Big ] \Big \} , 
\e
so we get,
\be
\int_0^1 g_T (x) dx = {e_q^2\over {2S^1}}\langle P, S^1 \mid M{\cal J}^1_{II} \mid P, S^1\rangle
\e
which explicitly shows the connection between the integral of the transverse
polarized structure function and the matrix element of the fermion
intrinsic-like part of
the transverse spin operator. 

It is quite instructive to compare our calculation of the transverse spin of
the dressed quark with the helicity of the dressed quark\cite{hk} in
perturbative QCD. All the operators contributing to helicity are kinematical
(interaction independent) and hence all of them give rise to only diagonal
contributions. Further, in this calculation mass of the quark can be
completely ignored since they give rise to only power-suppressed
contribution. In the massless limit, helicity is conserved at the quark
gluon vertex. This means that the quark in the quark-gluon state has the
same helicity as the parent quark. 
Since the transverse gluon carry helicity $\pm 1$, we get a non-vanishing
contribution from the gluon intrinsic helicity operator. However, both the
quark and the gluon in the quark-gluon state have non-vanishing orbital
angular momentum due to transverse motion. Total helicity conservation
implies that orbital contribution has to cancel gluon intrinsic helicity
contribution. This is precisely what happens\cite{hk} and
we find that the total quark plus gluon
orbital part exactly canceled the intrinsic gluon contribution and the
overall contribution to the helicity is $\pm{1\over 2}$, which entirely
comes from the intrinsic part of the fermionic helicity operator.

In contrast, in the case of transverse spin operator, it 
has both interaction independent and interaction dependent parts. The latter
gives rise to off-diagonal matrix elements and they play a very important
role. Of special interest is the gluon intrinsic-like transverse spin
operator. This operator gives vanishing matrix elements for a free gluon.
However, since gluon in the quark-gluon state has intrinsic transverse
momentum, both diagonal and off-diagonal terms give rise to non-vanishing
contributions and we get a net non-vanishing matrix element for the gluon
intrinsic-like transverse spin operator. However, we find that 
contribution from this matrix element is completely canceled by that from the matrix elements of
orbital-like transverse spin operators. This is analogous to what
happens in the helicity case.  

In this section, the calculation of the matrix elements has been done 
in the frame with
$P^\perp = 0$. The complete calculation of the matrix element of the
transverse spin operator in an arbitrary reference frame is given in 
appendix D. 
It is clear from the expressions there that all the terms explicitly
dependent on $P^\perp$ get canceled in the expectation value of $M{\cal
J}^1$. The parts that remain after the cancellation of
the $P^\perp$ dependent terms are those given above.
In the above expressions, we have used the manifest Hermitian form of the operators. 
We again stress the fact that this manifest cancellation of contributions
from center of mass motion is
typical in light-front  field theory because the transverse boost
operators are kinematical. The situation in the equal time relativistic case
is completely different and there one cannot separate out the center of mass
motion from the internal motion in a straightforward way even in the 
free theory case\cite{os}
because of the complicated boost generators. Due to the manifest cancellation
of the center of mass momenta, ${\cal J}^i$ can truly be identified as the
transverse spin operator. 

\section{summary and conclusions}

In this paper, we have analyzed the transverse spin operators in QCD. In equal
time quantization, one encounters two major difficulties in the description
of the spin of a composite system in an arbitrary reference frame. They are
1) the complicated interaction dependence arising from dynamical boost
operators and 2) the
difficulty in the separation of center of mass motion from the internal
motion. Due to these severe difficulties, there have been hardly any attempt
to study spin operators of a moving composite system in the conventional
equal time formulation of quantum field theory. 

In light-front theory, on the other hand, the longitudinal spin
operator (light-front helicity) is interaction independent and the
interaction dependence of transverse spin operators arises solely from that
of transverse rotation operators. Moreover, in this case the separation of
center of mass motion from internal motion is trivial since light-front
transverse boosts are simple Galilean boosts.

It is extremely interesting to contrast the 
cases of longitudinal and transverse spin operators
in light-front field theory. In the case of longitudinal spin operator
(light-front helicity), in the gauge fixed theory, the operator is
interaction independent and can be separated into orbital and spin parts for
quarks and gluons. It is known for a long time that the transverse spin 
operators in
light-front field theory cannot be separated into orbital and spin parts
except in the trivial case of free field theory.
{\it In this work, we have shown that, in spite of the complexities, 
a physically interesting separation is indeed
possible for the transverse spin operators} which is quite different from
the separation into orbital and spin parts in the rest frame familiar in the
equal time picture. We have discussed    
the physical significance of this separation.
Also  transverse rotational symmetry is not manifest in light-front theory 
and a study of these operators is needed for questions regarding Lorentz invariance
in the theory \cite{glazek}.

In analogy with the helicity sum rule, we have proposed a 
decomposition for the transverse spin. Elsewhere we have shown\cite{let}
the relationship between 
nucleon matrix elements of ${\cal J}^i_{II}$  and ${\cal J}^i_{III}$
and  
the first moments of quark and gluon structure functions respectively,
appearing in
transverse polarized hard scattering. This is the first time that
the integral of $g_T$ is related to a conserved quantity, namely the
transverse spin operator. It is important to mention here that the proposed
decomposition of the transverse spin operator will not be affected if one
adds a total derivative term to the angular momentum density. Such a term can
at most produce a surface term which we are neglecting since we have
restricted ourselves to the topologically trivial sector of the theory.
We have started with the angular momentum density defined in terms of the
symmetric gauge invariant stress-energy tensor, which is obtained from the
Noether's stress-energy tensor by a adding a total derivative term. 
Even though the angular momentum density differs from the Noether angular
momentum density by a total
derivative term, both give rise to the same generators.
Another point worth mentioning is that we have worked in the gauge fixed
theory. In the light-front gauge, $A^+=0$, the transverse spin operator can
be separated into three parts, and ${\cal J}_{II}^i$ is related to
the first moment of $g_T$ measured in transverse polarized scattering, which
is a gauge invariant object. This is similar to the helicity case, where
only in the light-front gauge and using light-front quantization, the intrinsic fermionic
helicity is related to  the gauge invariant first moment of $g_1$ measured in
longitudinally polarized scattering. The corresponding gluon intrinsic
helicity cannot be
measured directly in polarized deep inelastic lepton-nucleon scattering but in some other process
like polarized hadron-hadron scattering. A similar situation holds in the
case of transverse spin.

A very important issue related to the transverse
spin operators is renormalization. Because of the interaction dependence,
the operators acquire divergences in perturbation theory just like the
Hamiltonian and therefore have
to be renormalized. The renormalization of only the intrinsic-like fermion 
part of the 
transverse spin operator has been discussed in the literature so far. 
In this paper, we have carried out the renormalization of the
full transverse spin operator for the first time upto $O(\alpha_s)$ in 
light-front Hamiltonian perturbation theory by evaluating the matrix
elements for a dressed quark target.  We have shown that the entire
contribution to the matrix element comes from the fermion intrinsic-like 
part of the
transverse spin operator and is equal to ${1\over 2}$. The 
contributions from ${\cal J}^i_{I}$ and 
${\cal J}^i_{III}$  exactly get canceled. Also, the mass of the quark is
very crucial in this case, since the helicity flip interactions which are
proportional to the quark mass play a very important role.
However, the terms proportional to $m^2$ do not flip the helicity and do not
contribute. Since helicity flip is involved, we do not encounter any
quadratic divergence unlike the case of renormalization of the
light-front Hamiltonian.
Further, we have compared and contrasted the calculations of transverse spin
and helicity of a dressed quark in perturbation theory.       

We have also verified the frame independence of our results. We have
explicitly shown that, 
in an arbitrary
reference frame, all the terms
depending on the center of mass momenta manifestly get canceled in the
matrix element. The cancellation is as simple as in non-relativistic theory
since boost is kinematical on the light-front. For future studies, it is an
interesting problem to evaluate non-pertubatively \cite{bk} the matrix
element of the transverse spin operator in light-front QCD. Also, in this work,
we have used cutoff on the relative transverse momenta and the small $x$
divergence gets canceled in the one loop result. The renormalization of
the transverse spin operators using similarity renormalization technique
 \cite{wilson} is to be done in future.

\acknowledgments

We thank the organizers of the Workshop on High Energy Physics Phenomenology-VI,
Chennai, India, January 3-15, 2000 for warm hospitality where part of this
work was done. We acknowledge helpful discussions with Prof. P. P.
Divakaran. RR gratefully acknowledges the financial assistance of the
council of Scientific and Industrial Research (CSIR), India.

\appendix

\section{Poincare generators in light-front QCD}
In this appendix we derive the manifestly Hermitian kinematical Poincare 
generators (except
$J^3$) and the Hamiltonian in light-front QCD
starting from the gauge invariant symmetric energy momentum tensor
$\Theta^{\mu \nu}$. To begin with, $\Theta^{\mu \nu}$ is interaction
dependent. In the {\it gauge fixed} theory we find that the seven kinematical
generators are manifestly independent of the interaction.

We shall work in the gauge $A^+=0$ and ignore all surface terms. Thus we are
working in the completely gauge fixed sector of the theory\cite{hk}. The
explicit form of the operator $J^3$ in this case is given in 
Ref. \cite{hk} which is manifestly free of interaction at the operator level.
The rotation operators are given in sec. II.

At $x^+=0$, the operators $K^3$ and $E^i$ depend only on the density
$\Theta^{++}$. A straightforward calculation leads to 
\begin{eqnarray}
\Theta^{++} =  {\psi^+}^\dagger \st{\lh}{i\pp^+}\psi^+ + 
\partial^+ A^i \partial^+ A^i.
\end{eqnarray}
Then, longitudinal momentum operator,
\begin{eqnarray}
P^+ && = { 1 \over 2} \int dx^- d^2 x^\perp \Theta^{++} \nonumber \\
&& = { 1 \over 2} \int dx^- d^2 x^\perp \left [ {\psi^+}^\dagger 
\st{\lh}{i\pp^+} \psi^+ + \partial^+ A^j \partial^+ A^j \right ].
\end{eqnarray}
Generator of longitudinal scaling,
\begin{eqnarray}
K^3 && = - { 1 \over 4} \int dx^- d^2 x^\perp x ^- \Theta^{++}, \nonumber \\
&&=  - { 1 \over 4} \int dx^- d^2 x^\perp x ^- \left [
 {\psi^+}^\dagger \st{\lh}{i\pp^+}\psi^+ + \partial^+ A^j \partial^+ A^j
\right ].
\end{eqnarray}
Transverse boost generators,
\begin{eqnarray}
E^i && =  - { 1 \over 2} \int dx^- d^2 x^\perp x^i \Theta^{++}, \nonumber \\
&& = - { 1 \over 2} \int dx^- d^2 x^\perp x^i \left [  {\psi^+}^\dagger 
\st{\lh}{i\pp^+} \psi^+ + \partial^+ A^j \partial^+ A^j \right ].
\end{eqnarray}
The transverse momentum operator 
\begin{eqnarray}
P^i = { 1 \over 2} \int dx^- d^2 x^\perp \Theta^{+i}
\end{eqnarray}
which appears to have explicit interaction dependence.
Using the constraint equations for $\psi^-$ and $A^-$, we  have
\begin{eqnarray}
\Theta^{+i} && = \Theta^{+i}_F + \Theta^{+i}_G, \nonumber \\
\Theta^{+i}_F && = 2 {\psi^+}^\dagger i \partial^i \psi^+ + 2 g
{\psi^+}^\dagger A^i \psi^+, \\
\Theta^{+i}_G && = \partial^+ A^j \partial^i A^j - \partial^+ A^j \partial^j A^i
+ \partial^+ A^i \partial^j A^j - 2 g {\psi^+}^\dagger A^i \psi^+.
\end{eqnarray}
Thus 
\begin{eqnarray} 
P^i = { 1 \over 2} \int dx^- d^2 x^\perp \left [ 
 {\psi^+}^\dagger \st{\lh}{i\pp^i} \psi^+ +
  A^j \partial^+\partial^j A^i - A^i \partial^+\partial^j A^j 
- A^j\partial^+ \partial^i A^j  \right ].
\end{eqnarray}
Thus we indeed verify that all the kinematical operators are explicitly
independent of interactions. 

Lastly, the Hamiltonian operator can be written in the manifestly Hermitian
form as,
\begin{eqnarray}
P^- = {1 \over 2} \int dx^- d^2 x^\perp \Theta^{+-}
= {1\over 2}\int dx^-d^2x^\perp ( {\cal H}_0 + {\cal H}_{int}) 
\end{eqnarray}

where ${\cal H}_0$ is the free part given by,
\be
{\cal H}_0 =- A^j_a {(\pp^i)}^2 A^j_a + \xi^\dagger \Big [ {{-(\pp^\p)^2 +
m^2}\over {i\pp^+}}\Big ] \xi- \Big [ {{-(\pp^\p)^2 +
m^2}\over {i\pp^+}}\xi^\dagger\Big ] \xi .
\e   
The interaction terms are given by,
\be
{\cal H}_{int} = {\cal H}_{qqg} + {\cal H}_{ggg} + {\cal H}_{qqgg}+ 
{\cal H}_{qqqq} + {\cal H}_{gggg} ,
\e
where,
\be
 {\cal H}_{qqg} = -4g \xi^\dagger {1\over \pp^+} (\pp^\p \cdot A^\p)\xi
 +  g{\pp^\p\over {\pp^+}} [ \xi^\dagger ({\tilde \s}^\p \cdot A^\p) ] {\tilde \s}^\p
\xi + g\xi^\dagger ({\tilde \s}^\p \cdot A^\p) {1\over {\pp^+}} ({\tilde
\s}^\p \cdot \pp^\p) \xi \nonumber\\ ~~~~~ + g({ \pp^\p \over {\pp^+}}
\xi^\dagger) {\tilde \s}^\p ( { \tilde \s}^\p \cdot A^\p) \xi + g\xi^\dagger
{1\over {\pp^+}} ({\tilde \s}^\p \cdot  \pp^\p) ( {\tilde \s}^\p \cdot
A^\p) \xi \nonumber\\ ~~~~~~ -mg {1\over {\pp^+}} [ \xi^\dagger ({\tilde
\s}^\p \cdot A^\p) ] \xi + m g\xi^\dagger ( {\tilde \s}^\p \cdot A^\p){1\over
{\pp^+}} \xi
\nonumber\\~~~~~~~~~~ + mg ( {1\over {\pp^+}} \xi^\dagger) ({\tilde \s}^\p
\cdot A^\p) \xi - mg \xi^\dagger {1\over {\pp^+}} [( {\tilde \s}^\p \cdot
A^\p) \xi] ,
\e
\be
{\cal H}_{ggg} = 2gf^{abc} \Big [ \pp^iA^j_aA^i_bA^j_c + (\pp^i
A^i_a){1\over {\pp^+}}(A^j_b\pp^+A^j_c)\Big ] ,
\e
\be
{\cal H}_{qqgg}&& = g^2 \Big [ \xi^\dagger ({\tilde \s}^\p \cdot A^\p){1\over
{i\pp^+}} ({\tilde \s}^\p \cdot A^\p) \xi - {1\over {i\pp^+}} (\xi^\dagger {\tilde
\s}^\p \cdot A^\p) {\tilde \s}^\p \cdot A^\p \xi
\nonumber\\&&~~~+ 4{1\over \pp^+} (f^{abc} A^i_b\pp^+A^i_c){1\over \pp^+}
(\xi^\dagger T^a \xi)\Big ] ,
\e 
\be
{\cal H}_{qqqq} = 4g^2 {1\over \pp^+} (\xi^\dagger T^a \xi){1\over \pp^+}
(\xi^\dagger T^a \xi) ,
\e
\be
{\cal H}_{gggg} = &&{g^2\over 2} f^{abc}f^{ade} \Big [ A^i_bA^j_cA^i_dA^j_e
\nonumber\\&&~~~~~~~~~+ 2{1\over \pp^+} ( A^i_b\pp^+A^i_c){1\over
\pp^+}(A^j_d \pp^+A^j_e)\Big ] .
\e

\section{Transverse Spin for a System of Two Non-interacting Fermions}

In order to show the non-triviality of the transverse spin operators even in the free
theory and the manifest cancellation of the center of mass motion in this
case, here we
 evaluate the transverse spin for a composite system of two free
fermions.The manifest cancellation of the center of mass motion for the
interacting theory is much more complicated and is given in
appendix D.  

Let the mass of each fermion be $m$ and momenta $(k^+_i, k^\p_i)$, $ i=1,2$.
We take the state to be $\mid P \rangle = b^\dagger(k_1,s_1)b^\dagger(k_2,
s_2)\mid 0 \rangle $, where $s_1$ and $s_2$ are the helicities. 
\be
M{\cal J}^1
\mid P \rangle &&= \left ({ 1 \over 2} F^2 P^+ +K^3P^2 
 - { 1 \over 2} E^2 P^- - P^1
{\cal J}^3 \right ) 
\mid P \rangle .
\e

We introduce Jacobi momenta, $(x_i,q^\p)$ defined as,
\be
k_1^\p = q^\p + x_1P^\p, ~~k_2^\p = -q^\p + x_2P^\p~~~~~~~~~ k_i^+ = x_iP^+
\e
with $
\sum x_i = 1$.

Here $M$ is the mass of the composite system and $(P^+, P^\p)$ are the
momenta of the center of mass.

The partial derivatives with respect to the particle momenta can be
expressed in terms of these variables as,
\be
{\pp \over \pp k_1^i} = x_2{\pp \over \pp q^i} + {\pp \over \pp P^i},
~~~~~~~~~~~~~~~~~~{\pp \over \pp k_2^i} =  -x_1{\pp \over \pp q^i}
 + {\pp \over \pp P^i},
\e
and
\be
{\pp \over \pp k_1^+} = {x_2\over P^+}{\pp \over \pp x_1}+ {\pp \over \pp P^+}
- x_2{P^\p \over P^+}\cdot{\pp \over \pp q^\p}~, 
\e
\be
{\pp \over \pp k_2^+} = {x_1\over P^+}{\pp \over \pp x_2}+ {\pp \over \pp P^+}
+ x_1{P^\p \over P^+}\cdot{\pp \over \pp q^\p} ~.
\e  
Then we have,
\be
K^3P^2\mid P \rangle = [ -iP^2 x_1x_2({\pp \over \pp x_1} + {\pp \over \pp
x_2} ) -i P^2P^+ {\pp \over \pp P^+} ]\mid P \rangle ~,
\e 
\be
-{1\over 2}E^2P^-\mid P \rangle = \Big [-{i\over 2}( (P^\p)^2 + M^2 ) {\pp \over
\pp P^2} +iP^2 \Big ]\mid P \rangle~,
\e
\be
P^1{\cal J}^3 \mid P \rangle = [-i P^1(q^2{\pp \over \pp q^1} - 
q^1{\pp \over \pp q^2}) + P^1{s_1\over 2}+ P^1{s_2\over 2}]\mid P \rangle ~.
\e
\be
{1\over 2}F^2P^+ \mid P \rangle &&= \Big [iq^2 ( x_2{\pp \over \pp x_1} -
x_1{ \pp
\over \pp x_2})+ {s_1\over 2}{q^1\over x_1}
- {s_2\over 2}{q^1\over x_2}+ {i\over 2} ( m^2 + (q^\p)^2) ({x_2\over x_1} 
- {x_1\over x_2}) {\pp \over
\pp q^2}\nonumber\\&&~~~~+ {m\over 2} \sum_\lambda ( {\s^1_{\lambda s_1}\over
x_1} + {\s^1_{\lambda s_2}\over x_2}) 
 - iq^2P^\p \cdot{\pp \over \pp q^\p} +iP^2  x_1x_2({\pp \over \pp x_1}
+ {\pp \over \pp x_2}) \nonumber\\&&~~~~~~~+iP^+ P^2 {\pp \over \pp P^+} +
+{i\over 2}( m^2 + (q^\p)^2) {1\over {x_1 x_2}}
 {\pp \over \pp P^2} + {i\over 2} (P^\p)^2 {\pp \over \pp P^2}
\nonumber\\&&~~~~~~~~~+ i (
q^\p \cdot P^\p){\pp \over \pp q^2} 
~~  +{P^1\over 2}(s_1 + s_2)-iP^2 \Big ] \mid P \rangle ~ .
\e
Substituting
\be
M^2 = {( m^2 + (q^\p)^2)\over x_1x_2}~, 
\e
we get,
\be
M{\cal J}^1\mid P \rangle = &&\Big [iq^2 (x_2{\pp \over \pp x_1} - x_1{\pp \over \pp x_2}) + {i\over 2}(
m^2 + (q^\p)^2) ({x_2\over x_1} - {x_1\over x_2}) {\pp \over \pp q^2}
\nonumber\\&&~~~~ +{q^1\over 2}({s_1\over x_1} - {s_2\over x_2}) + {m\over
2}\sum_\lambda ({\s^1_{\lambda s_1}\over x_1}+{\s^1_{\lambda s_2}
\over x_2})\Big ] \mid P \rangle ~ .
\e 
Explicitly we see that $M{\cal J}^1$ does not depend on the center of mass 
momenta.

\section{Transverse Rotation Operator in QCD}

In this section we explicitly derive the expressions for interaction dependent
transverse rotation operators in light-front QCD starting from the
manifestly gauge invariant energy momentum tensor. 

We set the gauge $A^+=0$
and eliminate the dependent variables $\psi^-$ and $A^-$ using the equations
of constraint. In this paper we restrict to the topologically trivial sector
of the theory and set the boundary condition $A^i(x^-, x^i) \rightarrow 0 $
as $ x^{-,i} \rightarrow \infty$. This completely fixes the gauge and put
all surface terms to zero.

The transverse rotation operator 
\begin{eqnarray}
F^i = {1 \over 2} \int dx^- d^2 x^\perp \Big [ x^- \Theta^{+i} - x^i
\Theta^{+-} \Big ].
\end{eqnarray}
The symmetric gauge invariant energy momentum tensor 
\begin{eqnarray}
\Theta^{\mu \nu} &&= { 1 \over 2} {\overline \psi} \Big [ 
  \gamma^\mu i D^\nu + \gamma^\nu i D^\mu \Big ] \psi - F^{\mu \lambda a}
F^{\nu a}_{\, \, \lambda} 
 - g^{\mu \nu} \Big [ - { 1 \over 4} (F_{\lambda \sigma a})^2 +
{\overline \psi} ( \gamma^\lambda i D_\lambda - m) \psi \Big ],
\end{eqnarray}
where 
\begin{eqnarray}
i D^\mu &&= {1 \over 2} \st{\lh}{i\pp^\mu} + g A^\mu, \nonumber \\
F^{\mu \lambda a} && = \partial^\mu A^{\lambda a} - \partial^\lambda A^{\mu
a} + g f^{abc} A^{\mu b} A^{\lambda c}, \nonumber \\
F^{\nu a}_{\, \, \lambda} && = \partial^\nu A_{\lambda}^a - \partial_\lambda
A^{\nu a} + g f^{abc} A^{\nu b} A_\lambda^c.
\end{eqnarray}
First consider the fermionic part of $ \Theta^{\mu \nu}$:
\begin{eqnarray}
\Theta^{\mu \nu}_F = { 1 \over 2} {\overline \psi} \Big [ \gamma^\mu i D^\nu
+ \gamma^\nu i D^\mu \Big ]\psi - g^{\mu \nu } {\overline \psi} (\gamma^\lambda
i D_\lambda - m)\psi.
\end{eqnarray}
The coefficient of $g^{\mu \nu}$ vanishes because of the equation of motion. 

Explicitly, the contribution to $F^2$ from the fermionic part of
$\Theta^{\mu \nu}$ is given by
\begin{eqnarray}
F^2_F && = { 1 \over 2} \int dx^- d^2 x^\perp \left [ x^- { 1 \over 2}
{\overline \psi} (\gamma^+ i D^2 + \gamma^2 i D^+) \psi 
 - x^2 { 1 \over 2}{\overline \psi} (\gamma^+ i D^- + \gamma^- i D^+) \psi
\right ],  \nonumber \\
&& = F^2_{F(I)} + F^2_{F(II)},
\end{eqnarray} 
where 
\begin{eqnarray}
F^2_{F(I)}= { 1 \over 2} \int dx^- d^2 x^\perp x^- \Big [ 
{\psi^+}^\dagger {1 \over 2} \st{\lh}{i\pp^2} \psi^+ + {\psi^+}^\dagger g A^2 \psi^+ + 
{ 1 \over 4} {\overline
\psi} \gamma^i  \st{\lh}{i\pp^+}
\psi \Big ],
\end{eqnarray}
\begin{eqnarray}
F^2_{F(II)}= -{ 1 \over 2} \int dx^- d^2 x^\perp x^2 \Big [ 
{\psi^+}^\dagger \Big ({1 \over 2} \st{\lh}{i\pp^-} + gA^- \Big ) \psi^+ 
 + 
{ 1 \over 4}
{\psi^-}^\dagger \gamma^i \st{\lh}{i\pp^+}
\psi^- \Big ].
\end{eqnarray} 
We have the equation of constraint
\begin{eqnarray}
i \partial^+ \psi^- = \big [ \alpha^\perp \cdot ( i \partial^\perp + g
A^\perp) + \gamma^0 m \big ] \psi^+, \label{eoc}
\end{eqnarray}
and the equation of motion
\begin{eqnarray}
i \partial^- \psi^+ = -g A^- \psi^+ + \big [ \alpha^\perp \cdot  (i
\partial^\perp + g A^\perp) + \gamma^0 m \big]{ 1 \over i \partial^+}
\big [ \alpha^\perp \cdot  (i
\partial^\perp + g A^\perp) + \gamma^0 m \big]   \psi^+. \label{eom}
\end{eqnarray}
Using the Eqs. (\ref{eoc}) and (\ref{eom}) we arrive at free ($g$
independent) and
interaction ($g$ dependent) parts of $F^2_F$.
The free part of $F^2_F$ is given by
\begin{eqnarray}
F^2_{F(free)} &&= { 1 \over 2} \int dx^- d^2 x^\perp \Bigg \{x^- \Bigg [
\xi^\dagger \Big [i \partial^2 \xi\Big] - 
\Big [i \partial^2 \xi^\dagger\Big ] \xi \Bigg ]  \nonumber \\
&&~~~~~~ - x^2 \Bigg [ \xi^\dagger \Big [{ - (\partial^\perp)^2 +m^2 \over i
\partial^+} \xi \Big ] -  \Big [{ - (\partial^\perp)^2 +m^2 \over i
\partial^+} \xi^\dagger\Big ] \xi \Bigg ] \nonumber \\
&& ~~~~~~+ \Bigg [ \xi^\dagger \Big [ \sigma^3 \partial^1 + i \partial^2
\Big]{ 1 \over
\partial^+} \xi + \Big [ { 1 \over \partial^+} (\partial^1 \xi^\dagger \sigma^3 -
i \partial^2 \xi^\dagger) \Big ] \xi \Bigg ] \nonumber \\ 
&&~~~~~~ + m \Bigg [ \xi^\dagger \Big [{ \sigma^1 \over i \partial^+} 
\xi\Big ] -
\Big [{ 1 \over i \partial^+} \xi^\dagger\sigma^1\Big ] \xi \Bigg ]
\Bigg \}.
\end{eqnarray}
We have introduced the two-component field $\xi$, 
\begin{eqnarray} \psi^+ = 
\left [ \begin{array}{c} \xi \\
                       0 \end{array} \right ].
\end{eqnarray} 
The interaction dependent part of $F^2_{F(I)}$ is 
\begin{eqnarray}
F^2_{F(I)int} && = g \int dx^- d^2 x^\perp x^- \xi^\dagger A^2 \xi \nonumber
\\
&& ~~~~+ { 1 \over 4} g \int dx^- d^2 x^\perp \Big [ \xi^\dagger { 1 \over
\partial^+}[(-i \sigma^3 A^1 + A^2)\xi] + { 1 \over \partial^+}
[ \xi^\dagger (i \sigma^3 A^1 + A^2)]\xi \Big ].
\end{eqnarray} 

The interaction dependent part of $F^2_{F(II)}$ is 
\begin{eqnarray}
F^2_{F(II)int} =  { 1 \over 4} g \int dx^- d^2 x^\perp \Big 
[ \xi^\dagger { 1 \over
\partial^+}[(-i \sigma^3 A^1 + A^2)\xi] + { 1 \over \partial^+}
[ \xi^\dagger (i \sigma^3 A^1 + A^2)]\xi \Big ] \nonumber \\
- { 1 \over 2} g \int dx^- d^2 x^\perp x^2 \Bigg [
{\pp^\p\over {\pp^+}} [ \xi^\dagger ({\tilde \s}^\p \cdot A^\p) ] {\tilde \s}^\p
\xi + \xi^\dagger ({\tilde \s}^\p \cdot A^\p) {1\over {\pp^+}} ({\tilde
\s}^\p \cdot \pp^\p) \xi \nonumber\\ ~~~~~ + ({ \pp^\p \over {\pp^+}}
\xi^\dagger) {\tilde \s}^\p ( { \tilde \s}^\p \cdot A^\p) \xi + \xi^\dagger
{1\over {\pp^+}} ({\tilde \s}^\p \cdot  \pp^\p) ( {\tilde \s}^\p \cdot
A^\p) \xi \nonumber\\ ~~~~~~ -m {1\over {\pp^+}} [ \xi^\dagger ({\tilde
\s}^\p \cdot A^\p) ] \xi + m \xi^\dagger ( {\tilde \s}^\p \cdot A^\p){1\over
{\pp^+}} \xi
\nonumber\\~~~~~~~~~~ + m ( {1\over {\pp^+}} \xi^\dagger) ({\tilde \s}^\p
\cdot A^\p) \xi - m \xi^\dagger {1\over {\pp^+}} [( {\tilde \s}^\p \cdot
A^\p) \xi] \Bigg ] \nonumber\\  
~~~~- {1 \over 2 } g^2 \int dx^- d^2 x^\perp x^2 \Bigg [ 
\xi^\dagger {\tilde \sigma}^\perp \cdot A^\perp  { 1 \over i \partial^+} 
{\tilde \sigma}^\perp \cdot (A^\perp \xi)
- { 1 \over i \partial^+} (\xi^\dagger {\tilde \sigma}^\perp \cdot A^\perp) 
{\tilde \sigma}^\perp \cdot A^\perp \xi \Bigg ].
\end{eqnarray} 
We have introduced $ {\tilde \sigma}^1 =  \sigma^2$ and $ {\tilde
\sigma}^2 = - \sigma^1$.

Next consider the gluonic part of the operator $F^2$:
\begin{eqnarray}
F^2_g = { 1 \over 2} \int dx^- d^2 x^\perp \Big [ x^- \Theta^{+2}_g - x^2
\Theta^{+-}_g \Big ],
\end{eqnarray} 
where
\begin{eqnarray} 
\Theta^{+2}_g &&= - F^{+ \lambda a} F^{2a}_{\, \, \lambda}, \nonumber \\
\Theta^{+-}_g &&= - F^{+\lambda a} F^{- a }_{\, \, \lambda} + { 1 \over 4}
g^{+-}(F_{\lambda \sigma a})^2.
\end{eqnarray}
Using the constraint equation
\begin{eqnarray}
{ 1 \over 2} \partial^+ A^{-a} = \partial^i A^{ia} + g f^{abc} { 1 \over
\partial^+}(A^{ib} \partial^+A^{ic}) + 2 g { 1 \over \partial^+} \Big (
\xi^\dagger T^a \xi \Big ),
\end{eqnarray}
we arrive at
\begin{eqnarray}
F^2_g = F^2_{g(free)} + F^2_{g(int)}
\end{eqnarray}
where
\begin{eqnarray}
F^2_{g(free)} &&= { 1 \over 2} \int dx^- d^2 x^\perp \Bigg \{ x^- \Big (
A^{ja}\partial^+\partial^j A^{2a} - A^{2a}\partial^+ \partial^j A^{ja}+
A^{ja}\partial^+\partial^2 A^{ja}\Big ) \nonumber \\
&& ~~~~~~~~~~~ -x^2  \Big ( A^{ka}(\partial^j)^2 A^{ka} \Big ) \Bigg
\}
\nonumber \\
&& ~~~~~~~~~~~~~~~~~~~- 2\int dx^- d^2 x^\perp A^{2a} \partial^1  A^{1a}.
\end{eqnarray}
The interaction part
\begin{eqnarray}
F^2_{g(int)} &&= { 1 \over 2} \int dx^- d^2 x^\perp x^- \Bigg \{ 
gf^{abc} \partial^+ A^{ia} A^{2b} A^{ic} \nonumber \\
&&~~~~ + g\Big (  f^{abc} { 1 \over \partial^+}(A^{ib} 
\partial^+ A^{ic}) + 2{
1 \over \partial^+} (\xi^\dagger T^a \xi) \Big ) \partial^+ A^{2a} \Bigg \}
\nonumber \\
&&~~ - {1 \over 2} \int dx^- d^2 x^\perp  x^2 \Bigg \{
2g f^{abc} \partial^i A^{ja} A^{ib} A^{jc} + {g^2 \over 2} f^{abc} f^{ade}
A^{ib} A^{jc} A^{id} A^{je} \nonumber \\
&& ~~~~ + 2g \partial^i A^{ia} { 1 \over \partial^+}
\Big ( f^{abc}  A^{jb} \partial^+ A^{jc} + 2 \xi^\dagger T^a \xi \Big )
\nonumber \\
&& ~~~~ + g^2 \Big ( f^{abc} { 1\over \partial^+} (A^{ib}\partial^{+} A^{ic})
+2 { 1 \over \partial^+} \xi^\dagger T^a \xi \Big )
\Big ( f^{ade} { 1\over \partial^+} (A^{jd}\partial^{+} A^{je})
+ 2{ 1 \over \partial^+} \xi^\dagger T^a \xi \Big ) \Bigg \}.  
\end{eqnarray}
So the full transverse rotation operator in QCD can be written as,
\begin{eqnarray}
F^2  = F^2_{I} + F^2_{II} + F^2_{III},
\end{eqnarray}
where the explicit forms of 
$F^2_I$, $F^2_{II}$ and $F^2_{III}$ have been given in section II.

\section{transverse spin of a dressed quark in an arbitrary reference frame}

We introduce a wave packet state
\be
\mid \psi_\sigma \rangle = {1\over 2} \int dP^+ d^2P^\perp f(P) \mid P,
\sigma \rangle  \label{wp}
\e
which is normalized as,
\be
\langle \psi_\sigma \mid \psi_{\sigma'} \rangle = \delta_{\sigma \sigma'}.
\e
Here $f(P)$ is a function of $P$, the exact form of which is not important.
Using Eq. (\ref{nor}) we get,
\be
{1\over 2}\int dP^+ d^2P^\perp f^*(P) f(P)(2\pi)^3 P^+ = 1.
\label{norm}
\e 
The expectation values of the various operators involved in the definition
of $M{\cal J}^i$ are given below.
It is to be noted that we have done the calculation in an
arbitrary reference frame, in order to show that the dependence on the total
center of mass momenta $(P^+, P^\perp)$ actually gets canceled in the
expectation value of $M{\cal J}^i$.

The matrix elements presented below have been evaluated between wave packet
states of different helicities, namely $\sigma$ and
$\sigma'$. Since the transversely polarized  state can be
expressed in terms of the longitudinally polarized (helicity) states by
Eq. (\ref{dr1}), the matrix elements of these operators between transversely
polarized states can be easily obtained from these expressions.   
We introduce,
\be
\psi^\sigma_1 = f(P)\Phi^\sigma_1, ~~~~~~~~~~~~~~\psi^\sigma_{\sigma_1
\lambda} = f(P)\Phi^\sigma_{\sigma_1\lambda}.
\e
The matrix elements are given by, 
\be
\langle \psi_\sigma \mid K^3 P^2 \mid \psi_{\sigma' }\rangle&& =
{1\over 2} \int dP^+d^2P^\perp (2\pi)^3 P^+\Big ({i\over 2}\psi^{*\sigma}_1 
{\pp \psi^{\sigma'}_1 \over
{\pp P^+}} P^+ P^2  \nonumber\\&&~~~~
 + {i\over 2}\sum_{spin} \int dxd^2q^\perp P^2 
\psi^{\sigma*}_{\sigma_1 \lambda} {\pp \psi^{\sigma'}_{\sigma_1'
\lambda'}\over {\pp P^+}}P^+ + h. c.\Big ).
\e   
\be
\langle \psi_\sigma \mid {1\over 2}E^2P_{free}^- \mid \psi_{\sigma'} \rangle&& = 
{1\over 2} \int dP^+d^2P^\perp (2\pi)^3 P^+\Big (-{i\over
4}\psi^{*\sigma}_1 {\pp \psi^{\sigma'}_1\over {\pp P^2}}P^+{ (P^\perp)^2 +
m^2\over P^+}\nonumber\\&&~~~~~~~~~~~~~-{i\over 4}\sum_{spin}
 \int dxd^2q^\perp \psi^{*\sigma}_{\sigma_1 \lambda} {\pp \psi^{\sigma'}_{
\sigma'_1 \lambda'} \over {\pp P^2}}
 \Big [ {m^2 + (q^\perp+xP^\perp)^2\over x} \nonumber\\&&~~~~~~~~~~~~~~~~~~~~~~~~~~~~
+ {(-q^\perp + (1-x)P^\perp )^2 \over {1-x}}\Big ]+ h.c.\Big ).
\e
\be
\langle \psi_\sigma \mid {1\over 2}E^2P_{int}^- \mid \psi_{\sigma'} \rangle&& 
= -{g\over \sqrt {2(2\pi)^3}}{1\over 2}\int dP^+ d^2P^\perp(2\pi)^3 P^+
\sum_{spin} \int dx d^2q^\perp {1\over \sqrt {1-x}}\nonumber\\&&~~~~~
\Big (i\psi^{\sigma *}_1 {\pp \psi^{\sigma'}_{\sigma_1 \lambda}\over {\pp
P^2}}\chi^\dagger_{\sigma} ~\Big [ 
-{(q^\perp \cdot \epsilon^\perp) \over {1-x}} -{1\over 2} { (\tilde
\sigma^\perp \cdot \epsilon^\perp)(\tilde \sigma^\perp \cdot q^\perp)\over
x}\nonumber\\&&~~~~~~~~~~~~~~~~ -{1\over 2} im {(1-x)\over x}(\tilde 
\sigma^\perp \cdot \epsilon^\perp) \Big ] \chi_{\sigma_1} + h. c. \Big ).
\e
Here $h. c.$ is the Hermitian conjugate, $\sum_{spin}$ is summation over $\sigma_1, \sigma'_1, \lambda,
\lambda'$. $P^-_{free}$ is the free part and $P^-_{int}$ is the interaction part
of the light-front QCD Hamiltonian density.
\be
\langle \psi_\sigma \mid P^1 {\cal J}^3 \mid \psi{\sigma'} \rangle&& =
{1\over 2}\int dP^+ d^2P^\perp (2\pi)^3 P^+ \Big [\sum_{spin}
\int dxd^2q^\perp P^1 \Big ( {i\over 2}\psi^{*\sigma}_{\sigma_1 \lambda}
 ( q^2 {\pp \over{\pp q^1}} - q^1 {\pp \over {\pp q^2}} ) \psi^{\sigma'}_{
\sigma'_1 \lambda'}\nonumber\\&&+ h.c. \Big ) +{1\over 2} \int dxd^2q^\perp P^1 \sum_{
\lambda,\sigma_2,\sigma'_2}  \lambda \psi^{*\sigma}_{\lambda \sigma_2} 
\psi^{\sigma'}_{\lambda \sigma'_2} \nonumber\\&&~~~~~~~~~~~~~~~~~~~~~~~~
+ \int dxd^2q^\perp P^1 \sum_{\lambda,
\sigma_1, \sigma'_1}  \lambda \psi^{*\sigma}_{\sigma_1\lambda} 
\psi^{\sigma'}_{\sigma'_1 \lambda}\Big ] .
\e 
The first term in the above expression is the quark-gluon orbital part, the
second and the third terms are the intrinsic helicities of the quark and
gluon respectively.
Finally, the operator ${1\over 2}F^2P^+$ can be separated into three parts,
\be
{1\over 2}F^2P^+ = {1\over 2}F^2_IP^+ + {1\over 2}F^2_{II}P^+ +
 {1\over 2}F^2_{III}P^+
\e
where $F^2_I$,$F^2_{II}$ and $F^2_{III}$ have been defined earlier.
The matrix elements of the different parts of these operators for a dressed quark state
in an arbitrary reference frame are given below. A
part of this calculation has been given in some detail in appendix E.
\be
{1\over 2}F^2_IP^+ = {1\over 2}F^2_I(1)P^+ - {1\over 2}F^2_I(2)P^+ -
 {1\over 2}F^2_I(3)P^+.
\e
The matrix elements of these three parts are,
\be
\langle \psi_\sigma \mid {1\over 2}F^2_I(1)P^+ \mid \psi_{\sigma'} \rangle  &&=
\langle \psi_\sigma \mid {1\over 2} \int dx d^2q^\perp x^- P^2_0 {1\over
2}P^+ \mid \psi_{\sigma'} \rangle \nonumber\\&& = 
{1\over 2}\int dP^+ d^2P^\perp (2\pi)^3 P^+\Big [ -{i\over
2}\psi^{*\sigma}_1{\pp\psi^{\sigma'}_1\over {\pp P^+}}P^+
P^2\nonumber\\&&~~~~+{i\over 2}\sum_{spin}\int dxd^2q^\perp q^2 
\psi^{*\sigma}_{\sigma_1\lambda} P^\perp  {\pp 
\psi^{\sigma'}_{\sigma'_1\lambda'}\over {\pp q^\perp}}
\nonumber\\&&~~~~~~~~ -{i\over 2}\sum_{spin} \int dx d^2q^\perp  \psi^{*\sigma}_{
\sigma_1\lambda} P^2  {\pp \psi^{\sigma'}_{\sigma'_1\lambda'}\over {\pp
P^+}}P^+\nonumber\\&&~~~~~~~~~~~~~~ -{i\over 2}\sum_{spin} \int dx d^2
q^\perp q^2 \psi^{*\sigma}_{\sigma_1\lambda}{\pp\psi^{*\sigma'}_{\sigma'_1
\lambda'} \over {\pp x}} + h.c. \Big ].
\e
\be
\langle \psi_\sigma \mid && {1\over 2}F^2_I(2)P^+\mid \psi_{\sigma'} \rangle = 
\langle \psi_\sigma \mid {1\over 2} \int dx d^2q^\perp x^2 P^-_0 {1\over
2}P^+\mid \psi_{\sigma'} \rangle \nonumber\\&& = 
{1\over 2} \int dP^+ d^2P^\perp (2\pi)^3 P^+ \Big [{i\over
4}\psi^{*\sigma}_1 {\pp \psi^{\sigma'}_1\over {\pp P^2}}{ (P^\perp)^2 + m^2
\over P^+}P^+ \nonumber\\&&~~~~~+{i\over 4}\sum_{spin}
 \int dxd^2q^\perp \psi^{*\sigma}_{\sigma_1 \lambda} {\pp \psi^{\sigma'}_{
\sigma'_1 \lambda'} \over {\pp P^2}}
 \Big [ {m^2 + (q^\perp+xP^\perp)^2\over x} 
+ {(-q^\perp + (1-x)P^\perp )^2 \over {1-x}}\Big ]\nonumber\\&&~~~~~~~~+ 
{i\over 2}\sum_{spin} \int dx d^2q^\perp  \psi^{*\sigma}_{\sigma_1\lambda}{\pp \psi^{\sigma'}_{
\sigma'_1\lambda'}\over {\pp q^2}} (q^\perp \cdot P^\perp)
\nonumber\\&&~~~~~~~~~~~~~~~+{i\over 4}\sum_{spin}\int dxd^2q^\perp 
\psi^{*\sigma}_{\sigma_1\lambda}{\pp \psi^{\sigma'}_{\sigma'_1\lambda'}
\over {\pp q^2}} (q^\perp)^2 ( {{1-x}\over x} - {x\over {1-x}} ) 
\nonumber\\&&~~~~~~~~~~~~~~~~~~~~ +{i\over 4}\sum_{spin} \int dx d^2q^\perp m^2 {{1-x}\over x} \psi^{*\sigma}_{\sigma_1
\lambda}{\pp \psi^{\sigma'}_{\sigma'_1\lambda'}\over {\pp q^2}} 
 + h.c. \Big ].
\e
In the above two equations, both the single particle and two particle
diagonal matrix elements contribute.
\be
\langle \psi_\sigma \mid &&{1\over 2}F^2_I(3)P^+\mid \psi_{\sigma'} \rangle =
\langle \psi_\sigma \mid  {1\over 2} \int dx d^2q^\perp x^2 P^-_{int} {1\over
2}P^+\mid \psi_{\sigma'} \rangle \nonumber\\&& = {g\over \sqrt {2(2\pi)^3}}
\sum_{spin}{1\over 2}\int dP^+ d^2P^\perp (2\pi)^3P^+ \int dxd^2q^\perp {1\over {\sqrt {1-x}}} \Big ( i\psi^{\sigma *}_1 {\pp \psi^{\sigma'}_{\sigma_1 \lambda}\over {\pp
P^2}}\nonumber\\&&~~\chi^\dagger_{\sigma} \Big [ 
-{(q^\perp \cdot \epsilon^\perp) \over {1-x}} -{1\over 2} { (\tilde
\sigma^\perp \cdot \epsilon^\perp)(\tilde \sigma^\perp \cdot q^\perp)\over
x} -{1\over 2} im {(1-x)\over x}(\tilde 
\sigma^\perp \cdot \epsilon^\perp)
 \Big ]\chi_{\sigma_1}\nonumber\\&&~~~~~~ -{i\over 4} 
\psi^{* \sigma} \chi^\dagger_\sigma [ {\tilde \sigma}^2 ( {\tilde
\sigma}^\perp \cdot \ep^\perp )+ {({\tilde 
\sigma}^\perp \cdot \ep^\perp)
{\tilde \sigma}^2 \over x}]\chi_{\sigma_1} \psi^{\sigma'}_{\sigma_1
\lambda}+ h. c.\Big ).
\e
Only the off-diagonal matrix elements contribute in the above equation.
The matrix elements of the three different parts of ${1\over 2}F^2_{II}P^+$ are given by,
      
\be
\langle \psi_\sigma \mid {1\over 2}F^2_{mII}P^+ \mid \psi_{\sigma'}
\rangle&& =    
{1\over 2}\int dP^+ d^2P^\perp (2\pi)^3 P^+\Big [{m\over 2}\psi^{*\sigma}_1
\psi^{\sigma'}_1 \nonumber\\&&~~~~~~~~~+ {m\over 2}\sum_{spin} \int dx d^2q^\perp  \psi^{*\sigma}_{\sigma_1
\lambda} \chi^\dagger_{\sigma_1} \sigma^1 \chi_{\sigma'_1}
\psi^{\sigma'}_{\sigma'_1
\lambda'} {1\over x}\Big ],\label{b14}
\e
\be
\langle \psi_\sigma \mid {1\over 2}F^2_{q^\perp II} P^+ \mid \psi_{\sigma'}
 \rangle &&= {1\over 2}\int dP^+ d^2P^\perp (2\pi)^3 P^+\Big [
{1\over 2}\sum_{spin}\int dx d^2q^\perp \psi^{*\sigma}_{\sigma_1
\lambda} \chi^\dagger_{\sigma_1} \sigma^3 q^1 \chi_{\sigma'_1}
\psi^{\sigma'}_{\sigma'_1 \lambda'} {1\over x} \nonumber\\&&~~~~~~~~~~~~~
+ {1\over 2} \int dxd^2q^\perp \sum_{\lambda, \sigma_2, \sigma_2'} \lambda 
P^1\psi^{*\sigma}_{\lambda
\sigma_2} \psi^{\sigma'}_{\lambda \sigma_2'}\Big ],\label{b15}
\e
\be
\langle \psi_\sigma \mid {1\over 2}F^2_{g II} P^+ \mid \psi_{\sigma'}
 \rangle &&= {1\over 4} {g\over {\sqrt {2(2 \pi)^3}}}\sum_{spin} {1\over 2}
\int dP^+ d^2P^\perp (2\pi)^3 P^+\int dx d^2q^\perp
{1\over {\sqrt {1-x}}} \Big ( i\psi^{*\sigma}_1 \nonumber\\&&~~~~~~\Big [ 
\chi^\dagger_\sigma (
-i\sigma^3 \ep^1_\lambda + \ep^2_\lambda ) \chi_{\sigma_1} - {1\over x}
\chi^\dagger_\sigma (i\sigma^3 \ep^1_\lambda + \ep^2_\lambda ) \chi_{\sigma_1}
 \Big ] \psi^{\sigma'}_{\sigma_1 \lambda}  + h. c.\Big ).\label{b16}
\e
In Eqs. (\ref{b14}) and (\ref{b15}), contributions come from only diagonal
matrix elements whereas Eq. (\ref{b16}) contain only off-diagonal matrix
elements.
The matrix element of ${1\over 2}F^2_{III} P^+$ is given by,
\be
\langle \psi_\sigma \mid {1\over 2}F^2_{III} P^+&& \mid \psi_{\sigma'}
 \rangle = -{g \over {\sqrt {2(2\pi)^3}}}{1\over 2}
\int dP^+ d^2P^\perp (2\pi)^3 P^+\Big [\sum_{spin} \int dx d^2q^\perp {1\over
{\sqrt {1-x}}}\nonumber\\&&~~~ \Big (  \psi^{*\sigma}_1 \ep^2_\lambda 
\psi^{\sigma'}_{\sigma_1 \lambda}{1\over {i(1-x)}} + h. c. \Big )
 - \int dx d^2q^\perp {q^1\over (1-x)} \sum_{\lambda,
\sigma_1, \sigma'_1} \lambda
\psi^{*\sigma}_{\sigma_1 \lambda}\psi^{\sigma'}_{\sigma'_1
\lambda}\nonumber\\&&~~~~~~~~~~~~~~~~~~~~~~~~~~~~~~~~+ 
\int dx d^2q^\perp P^1 \sum_{\lambda, \sigma_1, {\sigma'}_1} \lambda 
\psi^{*\sigma}_{\sigma_1 \lambda} 
\psi^{\sigma'}_{\sigma'_1 \lambda}\Big ] . 
\e

Finally, the expectation value of the transverse spin operator is given by,
\be
\langle \psi_{s^1} \mid M{\cal J}^1 \mid \psi_{s^1}
 \rangle = \langle \psi_{s^1} \mid {1\over 2} F^2 P^+ + K^3P^2 - {1\over 2}
E^2 P^- - P^1 {\cal J}^3 \mid \psi_{s^1} \rangle .
\e
From the above expressions it is clear that all the explicit $P^\perp$
dependent terms get canceled in the final expression. 
To be specific, it can be easily seen that all the terms in the expectation
value of $K^3P^2-{1\over 2}E^2P^-_{free} - P^1{\cal J}^3_{orbital}$ are
$P^\perp$ dependent and they exactly cancel the $P^\perp$ dependent terms in
${1\over 2} F^2_{I(free)}P^+$; the two $P^1$ dependent terms in the
intrinsic part of $P^1{\cal J}^3$ exactly cancel the two similar terms in
the expectation value of ${1\over 2}F^2_{II}P^+ + {1\over 2}F^2_{III}P^+$
and the expectation value of ${1\over 2} E^2 P^-_{int}$ completely cancel all
the $P^\perp$ dependent terms in the expectation value of ${1\over
2}F^2_{I(int)}P^+$.  
\section{ Details of the calculation}
Here, we explicitly show the evaluation of one of the matrix elements of the
interaction part of $F^2_I$.
Consider the operator,
\be
O_g = {1\over 2} \int dx^- d^2x^\perp x^2 g \Big [{( \tilde \sigma^\perp 
\cdot \pp^\perp + m)\over {\pp^+}}\xi^\dagger \Big ] ( \tilde \sigma \cdot
A^\perp) \xi {P^+\over 2}.
\e 
This can be written in Fock space as,
\be
O_g=&& {g\over 2}\sum_{s_1, s_2, \lambda}\int (dk_1) \int (dk_2) \int [dk_3] \Big ( b^\dagger ( k_1, s_1 )
a ( k_3, \lambda) b( k_2, s_2 ) \chi^\dagger_{s_1} {( \tilde \sigma^\perp 
\cdot k_1^\perp - im)\over k_1^+}( \tilde \sigma^\perp
\cdot \ep^\perp_\lambda)  \chi_{s_2}\nonumber\\&& i {\pp\over {\pp k_1^2}} 2(2\pi)^3 \delta^3(
k_1 - k_2 - k_3 ) + b^\dagger ( k_1, s_1 )
a^\dagger ( k_3, \lambda) b( k_2, s_2 ) \chi^\dagger_{s_1} 
{( \tilde \sigma^\perp 
\cdot k_1^\perp - im)\over k_1^+}( \tilde \sigma^\perp \cdot 
\ep^{*\perp}_\lambda)  \chi_{s_2}\nonumber\\&&~~~~~~~~~~~~~~~~~~~~~~~~~~~~~
i {\pp\over {\pp k_1^2}} 2(2\pi)^3 \delta^3(k_1 - k_2 + k_3 ) \Big ){P^+ 
\over 2}
\e
where $ (dk) = {dk^+ d^2k^\perp\over {2(2\pi)^3}\sqrt k^+}$ and $[dk] = {dk^+ d^2
k^\perp\over { 2(2\pi)^3 k^+}}$.
We evaluate the expectation value of this operator for the dressed quark
state given by Eq.(\ref{wp}). Only the off-diagonal parts of the matrix element will
give non-zero contribution. The matrix element is given by,
\be
\langle \psi_\sigma \mid O_g \mid \psi_{\sigma'} \rangle &&=
{g\over 2} \sum_{\sigma_1, \lambda}\int (dP)'\int \{ dp_1 \} \int \{ dp_2 \} \sqrt {2(2\pi)^3}
P^+\delta^3(P-p_1 - p_2)\nonumber\\&&~~~~~\{ \phi_1^{*\sigma}\sqrt {p^+_1}
 \chi^\dagger_\sigma {{(\tilde \sigma^\perp \cdot P^\perp -
im) }\over P^+}( \tilde \sigma^\perp \cdot \ep_\lambda^\perp) \chi_{\sigma_1}
\phi^{\sigma'}_{ \sigma_1 \lambda'} i{\pp\over \pp P^2}\nonumber\\&&~~~~~~~~~~~~~~~~~~~~~~
 2(2\pi)^3 \delta^3 (P-p_1-p_2) {1\over 2} P^+ + h. c. \}
\e  
where $\{dp \} = {dp^+ d^2p^\perp \over {\sqrt {2 (2\pi)^3 p^+}}}$ and $ (dP)'
= {1\over 2}dP^+ d^2p^\perp 2(2\pi)^3P^+$. 
\be
=-{ig\over 4}\int (dP)' \int \{ dp_1 \} &&\int \{ dp_2 \}\sum_{\sigma_1, \lambda} \sqrt {2(2\pi)^3} P^+ 
\Big [\phi_1^{*\sigma}[{\pp \over {\pp P^2}}\phi^{\sigma'}_{ 
\sigma_1 \lambda}\delta^3(
P-p_1 - p_2)\sqrt {p^+_1}\nonumber\\&&~~~~~~ \chi^\dagger_\sigma 
{{( \tilde \sigma^\perp \cdot P^\perp -im) }\over P^+}( \tilde 
\sigma^\perp \cdot \ep_\lambda^\perp)\chi_{\sigma'}\Big ]
  2(2\pi)^3 \delta^3 (P-p_1-p_2) P^++ h. c. \Big ]\nonumber
\e  
\be
=-{ig\over 4}{1\over {\sqrt {2(2\pi)^3}}}&&\sum_{\sigma_1, \lambda}\int (dP)' \int dx d^2 q^\perp
{1\over {\sqrt {1-x}}}\Big [ \psi_1^{*\sigma} {\pp \psi^{\sigma'}_{\sigma_1 
\lambda}\over {\pp
P^2}}\chi^\dagger_\sigma ( \tilde \sigma^\perp \cdot P^\perp - im)(\tilde 
\sigma^\perp \cdot \ep_\lambda^\perp)\chi_{\sigma'}\nonumber\\&&~~~~~
 + h. c. \Big ] -{ig\over 4}{1\over {\sqrt {2(2\pi)^3}}}\sum_{
\sigma_1, \lambda}\int (dP)' \int dx d^2 q^\perp{1\over {\sqrt
{1-x}}}\Big [ \psi_1^{*\sigma} \psi^{\sigma'}_{\sigma_1 \lambda}
\nonumber\\&&~~~~~~~~~~~~~~~~~~\chi^\dagger_{\sigma} {\tilde \sigma}^2 ( \tilde \sigma^\perp \cdot
\epsilon^\perp_\lambda ) \chi_{\sigma'} + h. c. \Big ]   
\e     
The other terms can also be evaluated in a similar method.


\eject
\end{document}